\PassOptionsToPackage{hypertexnames=false}{hyperref}

\documentclass[prl, reprint, superscriptaddress, floatfix, longbibliography]{revtex4-2}

\usepackage{mathtools}
\usepackage{textgreek}
\usepackage{graphicx}
\usepackage{bm}
\usepackage{hyperref}
\usepackage{braket}
\usepackage{xcolor}
\usepackage{multirow}
\usepackage{placeins}
\usepackage{mleftright}
\usepackage{makecell}
\newcommand{\ketbra}[2]{|#1\rangle\!\langle #2|}
\newcommand{\avg}[1]{\langle #1 \rangle}
\newcommand{\op}[1]{\hat{#1}}

\newcommand{\dn}{\delta \op{n}}
\newcommand{\OR}{\Omega_{\mathrm{q}}}

\newcommand{\Gm}{\Gamma_m}
\newcommand{\Grabi}{\Gamma_{\mathrm{Rabi}}}
\newcommand{\Gonerho}{\Gamma_{1\rho}}
\newcommand{\Gtworho}{\Gamma_{2\rho}}
\newcommand{\Gphirho}{\Gamma_{\varphi\rho}}
\newcommand{\nbar}{\bar{n}}

\newcommand{\gesubspace}{\{\ket{g},\,\ket{e}\}}

\begin{document}

\title{Heralded Leakage Detection with Preserved Computational-State Coherence in a Fixed-Frequency Transmon}

\author{Takeaki~Miyamura}
\email{miyamura@qipe.t.u-tokyo.ac.jp}
\affiliation{Department of Applied Physics, Graduate School of Engineering, The University of Tokyo, Bunkyo-ku, Tokyo 113-8656, Japan}
\author{Zhiling~Wang}
\affiliation{RIKEN Center for Quantum Computing (RQC), Wako, Saitama 351-0198, Japan}
\author{Peter~A.~Spring}
\affiliation{RIKEN Center for Quantum Computing (RQC), Wako, Saitama 351-0198, Japan}
\author{Hiroto~Mukai}
\affiliation{Department of Applied Physics, Graduate School of Engineering, The University of Tokyo, Bunkyo-ku, Tokyo 113-8656, Japan}
\author{Kohei~Matsuura}
\affiliation{Department of Applied Physics, Graduate School of Engineering, The University of Tokyo, Bunkyo-ku, Tokyo 113-8656, Japan}
\author{Jesper~Ilves}
\affiliation{Department of Applied Physics, Graduate School of Engineering, The University of Tokyo, Bunkyo-ku, Tokyo 113-8656, Japan}
\author{Yoshiki~Sunada}
\affiliation{Department of Applied Physics, Stanford University, Stanford, California 94305, USA}
\author{Yasunobu~Nakamura}
% \email{yasunobu@ap.t.u-tokyo.ac.jp}
\affiliation{Department of Applied Physics, Graduate School of Engineering, The University of Tokyo, Bunkyo-ku, Tokyo 113-8656, Japan}
\affiliation{RIKEN Center for Quantum Computing (RQC), Wako, Saitama 351-0198, Japan}

\date{\today}

\begin{abstract}
Leakage out of the computational subspace is a major error mechanism in superconducting quantum processors.
Detecting leakage without disturbing the encoded quantum information can provide a heralded error signal that error-correction decoders exploit.
However, standard dispersive readout collapses all qubit eigenstates indiscriminately.
Here, we demonstrate heralded leakage detection on a fixed-frequency transmon by applying a Rabi drive on the computational transition during dispersive readout.
The drive makes the computational states indistinguishable to the probe on the resonator while leaving the second and higher excited states distinguishable from the computational states.
We achieve a leakage-detection fidelity of 97.1(3)\% within the 80-ns detection window, with a false-flag rate of 2.3(3)\% from the computational subspace.
Conditioned on the no-leakage outcome, the post-detection state retains an average fidelity of 92.9(5)\% across the six cardinal states for an equal mixture of computational and leakage population.
The scheme requires no circuit elements beyond those used for standard dispersive readout, making it applicable to various types of superconducting qubits without hardware modification.
\end{abstract}

\maketitle

Superconducting quantum circuits are one of the leading platforms for quantum computation.
Quantum error correction on superconducting quantum processors has recently reached the surface-code threshold and beyond-break-even logical lifetimes~\cite{krinner_realizing_2022, google_quantum_ai_suppressing_2023, acharya_quantum_2025}, and further improvements in the control and readout of qubits are expected to accelerate this progress.

Superconducting qubits are usually realized as the two lowest energy levels of anharmonic multilevel systems, where unwanted population of higher levels constitutes leakage from the computational subspace.
Because leakage falls outside the Pauli-error model assumed by stabilizer codes, it can persist over many error-correction cycles and propagate through multi-qubit gates, generating correlated
errors that are harmful to code performance~\cite{varbanov_leakage_2020,miao_overcoming_2023}.
Detecting which qubits are in the leakage state, without disturbing the encoded quantum information, is therefore a key capability for leakage management.
Such heralded detection enables post-selective removal of leakage and supplies a flag to the decoder~\cite{stace_error_2010}, potentially complementing the unconditional leakage reduction units~\cite{battistel_hardware-efficient_2021,mcewen_removing_2021,marques_all-microwave_2023, yang_coupler-assisted_2024, lacroix_fast_2025, xin_improved_2025, camps_leakage_2026, martin-vazquez_passive_2025}.
Heralded leakage detection is also relevant to the dual-rail-encoded microwave-photonic quantum communication, where photon loss is mapped to the second-excited-state population of a receiving qubit~\cite{kurpiers_quantum_2019, ilves_-demand_2020, yang_deterministic_2025, wang_generation_2026}.

Several approaches to distinguishing leakage states from the computational subspace have been demonstrated~\cite{jerger_realization_2016, liu_converting_2026} or proposed~\cite{wang_proposal_2026} based on engineered dispersive shifts.
However, they rely on specific qubit symmetries or qubit--resonator parameters for their operation. On the other hand, inference from repeated stabilizer outcomes provides only indirect access and adds decoding complexity~\cite{bultink_protecting_2020, varbanov_leakage_2020}. A direct measurement that flags leakage without such hardware-level constraints, while preserving computational-subspace coherence, remains an open challenge.
 
Continuous spin locking of a qubit with a resonant Rabi drive is known to suppress low-frequency
dephasing~\cite{yan_rotating-frame_2013,yoshihara_flux_2014,gustavsson_dynamical_2012, laucht_dressed_2017, abdurakhimov_driven-state_2020, sung_multi-level_2021, salhov_optimally_2026}.
When applied during dispersive readout, the drive averages out the state-dependent cavity pull, strongly suppressing measurement-induced dephasing~\cite{smirnov_decoherence_2003,szombati_quantum_2020}.
 
In this Letter, we apply this mechanism to realize heralded leakage detection on a fixed-frequency transmon.
By applying a Rabi drive near the computational transition frequency while simultaneously probing the readout resonator, the computational subspace becomes transparent to the measurement, while the leakage states remain dispersively coupled to the resonator and are thereby flagged.
We achieve a leakage-detection fidelity of $97.1(3)\%$ with a detection window of $80~\mathrm{ns}$.
We further perform quantum state tomography after the detection on the initial states containing both computational and leakage populations.
For an equal mixture, the post-detection average state fidelity reaches $92.9(5)\%$.
The protocol requires no flux tunability, ancilla qubit, or additional microwave component, providing a hardware-efficient route to leakage detection in circuit-QED systems.

\begin{figure}[t]
    \centering
    \includegraphics{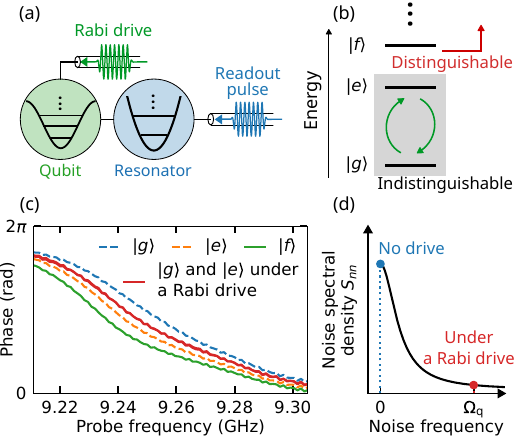}
    \caption{(a)~Scheme of the leakage-detection protocol. (b)~Energy-level diagram of the qubit. When the computational subspace is driven with sufficient power, the states $\ket{g}$ and $\ket{e}$ cannot be distinguished, while the higher states such as $\ket{f}$ can be detected through dispersive readout. (c)~Measured resonator phase response as a function of the readout frequency for $\ket{g}$~(blue dashed), $\ket{e}$~(orange dashed), $\ket{f}$~(green solid), and the computational subspace under the Rabi drive of strength~$\OR/2\pi=74\mathrm{\,MHz}$~(red solid). (d)~Illustration of the photon-number noise spectral density $S_{nn}(\omega)$. The blue (red) dot indicates the case without (with) a Rabi drive.}
    \label{fig:1}
\end{figure}

The detection protocol exploits the distinct responses of the computational subspace and the higher excited states under a near-resonant Rabi drive. 
Figure~\ref{fig:1}(a) shows the schematic of the system, where the qubit is dispersively coupled to a readout resonator.
The energy levels of the qubit are shown in Fig.~\ref{fig:1}(b), where $\ket{g}$, $\ket{e}$, and $\ket{f}$ denote the ground, first excited, and second excited states, respectively.
During dispersive readout, a Rabi drive nearly resonant with the $\ket{g}$--$\ket{e}$ transition frequency is applied.
The resulting Rabi oscillations time-average the state-dependent resonator pull, merging the resonator responses of $\ket{g}$ and $\ket{e}$ and rendering these two states indistinguishable~\cite{szombati_quantum_2020}.
Since $\ket{f}$ and higher states are not resonantly driven, their dispersive shifts remain intact, allowing them to be detected through the readout signal.
Figure~\ref{fig:1}(c) shows the measured resonator phase response, confirming that the $\ket{g}$ and $\ket{e}$ responses merge under the Rabi drive while the $\ket{f}$ response remains distinct.
We have verified that the $\ket{f}$-state phase response is unchanged under the Rabi drive.

Although the Rabi drive makes the computational states indistinguishable, the qubit still experiences dephasing induced by the readout photons.
In the frame of the Rabi drive, the computational subspace is described by two dressed states separated by the Rabi frequency. 
We denote the longitudinal and transverse relaxation rates of these dressed states by $\Gonerho$~and $\Gtworho$, respectively.
Both rates are affected by the noise of the resonator photon number through the dispersive coupling~$\chi\op{n}\ketbra{e}{e}$, where $\op{n}$ is the photon-number operator of the resonator and $\chi$ is the full dispersive shift.
Because this coupling is transverse with respect to the dressed states, the resulting relaxation samples the photon-number noise at the Rabi frequency rather than zero frequency.
The measurement-induced contribution to the longitudinal relaxation rate is
\begin{equation}\label{eq:G_meas_main}
  \Gonerho^{\mathrm{meas}} = \frac{\chi^2}{2} S_{nn}(\OR),
\end{equation}
where $\OR$ is the Rabi-drive strength and $S_{nn}(\omega)$ is the photon-number noise spectral density of the resonator.
See Sec.~II\,C of the Supplemental Material~\cite{SuppMat} for the derivation.
For a single resonator coupled to a feed line, the spectrum takes a Lorentzian form, 
\begin{equation}\label{eq:Snn_lorentz}
    S_{nn}(\omega) = \frac{\nbar\kappa}{\omega^2 + (\kappa/2)^2},
\end{equation}
where $\nbar$ is the mean intra-cavity photon number and $\kappa$ is the linewidth of the resonator. 
Without the Rabi drive~($\OR=0$), Eq.~\eqref{eq:G_meas_main} reduces to the standard measurement-induced dephasing rate~$\Gm=2\chi^2\nbar/\kappa$~\cite{gambetta_qubit-photon_2006, he_effect_2023}.
The Rabi drive shifts the noise sampling frequency from zero to $\OR$, and thus the measurement-induced dephasing is suppressed under the condition~$\OR \gg \kappa$~\cite{szombati_quantum_2020}~[Fig.~\ref{fig:1}(d)].

The above analysis treats the qubit as a two-level system. 
For a transmon with anharmonicity $\alpha/2\pi$ of typically $-200$ to $-300$~MHz, a strong Rabi drive ($\OR/2\pi\sim100$~MHz) mixes a small $\ket{f}$ component into the dressed computational states $\ket{\tilde 0}$ and $\ket{\tilde 1}$ through the virtual $\ket{e}$--$\ket{f}$ coupling. 
This admixture generates longitudinal noise in the dressed basis. 
This longitudinal noise samples $S_{nn}(0)$ rather than $S_{nn}(\OR)$ and is therefore not suppressed by the Rabi drive~(see Sec.~II\,D of the Supplemental Material).
However, the longitudinal noise can be canceled by introducing a small drive detuning $\delta\equiv\omega_{\mathrm{d}} -\omega_{eg}$ of the drive frequency $\omega_{\mathrm{d}}$ from the $\ket{g}$--$\ket{e}$ transition frequency $\omega_{eg}$, 
\begin{equation}\label{eq:delta_clock_main}
  \delta \approx \OR\left(-\frac{\OR}{2\alpha} + O\left((\OR/\alpha)^2\right)\right).
\end{equation}
To leading order in $|\OR/\alpha|$, this condition agrees with the spin-locking theory~\cite{zuk_robust_2024}. 

% \section{Experiment}
\begin{figure}
    \centering
    \includegraphics{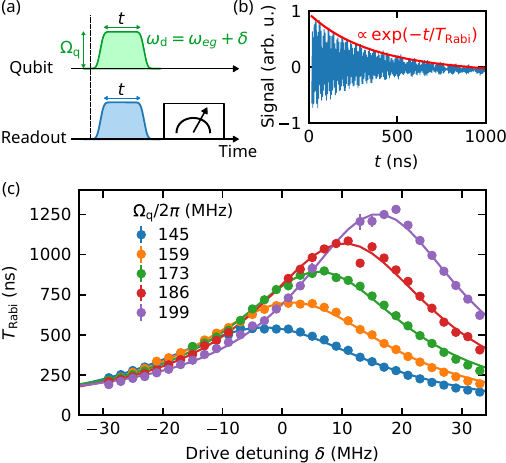}
    \caption{(a)~Pulse sequence to measure Rabi oscillations under a simultaneous probe pulse. A qubit drive at frequency $\omega_\mathrm{d} = \omega_{eg} + \delta$ and a readout pulse are applied concurrently for a duration~$t$, followed by a standard dispersive measurement. (b)~Observed Rabi oscillations with drive strength $\Omega_{\mathrm{q}}/2\pi=186$~MHz, drive detuning $\delta/2\pi=-20$~MHz, and mean photon number~$\nbar=2.41$. The red curve is the exponential fit of the envelope, from which the decay time $T_{\mathrm{Rabi}}$ is extracted. (c)~$T_{\mathrm{Rabi}}$ as a function of the drive detuning $\delta$ for different drive strengths $\Omega_{\mathrm{q}}$ at $\nbar=2.41$. Solid lines are the Lorentzian fits. Data points affected by a spurious mode are omitted.}
    \label{fig:2}
\end{figure}
We experimentally investigate the proposed leakage detection using the same device as in Refs.~\citenum{wang_generation_2026}~and \citenum{miyamura_generation_2025}. 
The device consists of a fixed-frequency transmon dispersively coupled to a readout resonator, which is coupled to a band-pass Purcell filter~(see Sec.~I of the Supplemental Material for the details).
An active reset is applied at the beginning of every pulse sequence throughout this work~\cite{magnard_fast_2018}.

We first characterize the suppression of measurement-induced dephasing.
As shown in Fig.~\ref{fig:2}(a), a Rabi drive at frequency $\omega_\mathrm{d}=\omega_{eg}+\delta$ and a probe tone on the readout resonator are applied simultaneously for a duration
$t$, followed by a standard dispersive readout.
The probe frequency is set to $9.2468$~GHz, the standard $\ket{g}$--$\ket{e}$ readout frequency.
Figure~\ref{fig:2}(b) shows a representative Rabi-oscillation trace at $\OR/2\pi=186$~MHz, $\delta/2\pi=-20$~MHz, and mean intra-cavity photon number $\nbar=2.41$, where $\nbar$ is averaged over the $\ket{g}$ and $\ket{e}$ states~(see Sec.~I of the Supplemental Material).
The oscillations are asymmetric about zero because the drive is detuned from the qubit transition frequency~(see Sec.~II\,F of the Supplemental Material).
The envelope decays with a characteristic time $T_{\mathrm{Rabi}}=1/\Gtworho$.
Accounting for the measurement-induced dephasing, we write the transverse relaxation rate as~\cite{yan_rotating-frame_2013}
\begin{equation}
    \Gtworho = \frac{3}{4}\Gamma_1 + \Gtworho^{\mathrm{meas}},
\end{equation}
where $\Gamma_1$ is the intrinsic energy-relaxation rate and $\Gtworho^{\mathrm{meas}}=\Gonerho^{\mathrm{meas}}/2$ is the measurement-induced transverse relaxation rate of the drive-dressed states~(see Sec.~II of the Supplemental Material).
At $\nbar=2.41$, the measured $T_{\mathrm{Rabi}}$ is of order $1~\mathrm{\mu s}$, much shorter than $T_1\approx27~\mathrm{\mu s}$, confirming that the decay is dominated by measurement-induced noise.

We repeat this measurement while sweeping $\OR$ and $\delta$, extracting $T_{\mathrm{Rabi}}$ by fitting the Rabi-oscillation trace~(see Sec.~II\,F of the Supplemental Material for the fitting function).
Figure~\ref{fig:2}(c) shows the extracted $T_{\mathrm{Rabi}}$ as a function of $\delta$ for several values of $\OR$ at $\nbar=2.41$.
For each $\OR$, $T_{\mathrm{Rabi}}$ exhibits a peak at a detuning that increases with $\OR$, consistent with Eq.~\eqref{eq:delta_clock_main}.
The peak value of $T_{\mathrm{Rabi}}$ also increases with $\OR$, reflecting the noise suppression predicted by Eqs.~\eqref{eq:G_meas_main} and \eqref{eq:Snn_lorentz}.
The optimal Rabi-drive frequency shifts downward as the probe power is increased, due to the ac Stark shift of the qubit frequency~(see Sec.~II\,D of the Supplemental Material).
From these measurements, we select the Rabi-drive parameters $\OR/2\pi=186$~MHz and $\delta/2\pi=+11$~MHz.
The probe power, corresponding to $\nbar=2.41$, is chosen as a compromise between the signal-to-noise ratio of the leakage detection and the measurement-induced decay of the computational subspace.
Although the peak $T_{\mathrm{Rabi}}$ keeps growing with $\OR$, we operate at $\OR/2\pi=186$~MHz, because the strongest drive induces a beating pattern in the Rabi-oscillation signal near its $T_{\mathrm{Rabi}}$ peak, which we attribute to a spurious mode near the qubit frequency~(see Sec.~II\,G of the Supplemental Material).
\begin{figure}
    \centering
    \includegraphics{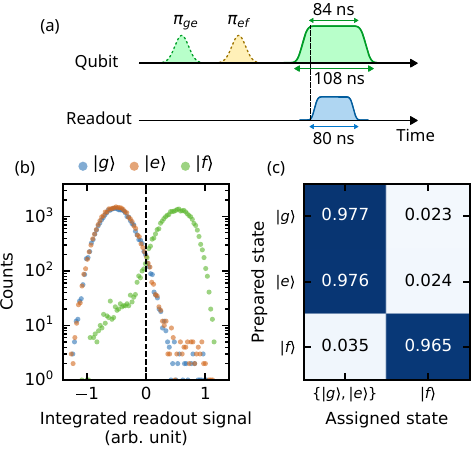}
    \caption{(a)~Pulse sequence for the leakage detection experiment. (b)~Single-shot readout histograms for prepared states $\ket{g}$~(blue), $\ket{e}$~(orange), and $\ket{f}$~(green). The dashed line indicates the classification threshold. (c)~Assignment matrix for binary \{$\ket{g},\,\ket{e}$\} or $\ket{f}$ classification.}
    \label{fig:3}
\end{figure}

With the optimized drive parameters, we characterize the single-shot leakage-detection readout.
The pulse sequence is shown in Fig.~\ref{fig:3}(a).
We prepare $\ket{g},\,\ket{e}$, or $\ket{f}$ using corresponding 40-ns-duration $\pi$ pulses and apply the Rabi drive and readout pulse simultaneously.
The Rabi drive (readout pulse) has a flat-top shape, with 12-ns (4-ns) Gaussian edges and total duration of 108~ns~(80~ns).
We use a flux-driven Josephson parametric amplifier~\cite{yamamoto_flux-driven_2008, kono_high-gain_2026} in the degenerate mode for improved measurement efficiency.
For leakage detection, the readout frequency is shifted by $-10$~MHz from the standard $\ket{g}$--$\ket{e}$ readout frequency.
This offset increases the signal contrast for the leakage states because the leakage states pull the resonator to a lower frequency.
It also reduces the photon number that the computational subspace experiences from $\nbar=2.41$ to $\nbar=2.28$.
The reflected signal is integrated over the readout window, and each single-shot record is classified into either $\{\ket{g},\,\ket{e}\}$ or $\{\ket{f}\}$.
Figure~\ref{fig:3}(b) displays the readout histograms with 30000~shots for each prepared state.
The $\ket{g}$ and $\ket{e}$ distributions overlap, confirming that the computational states are indistinguishable under the Rabi drive, while the $\ket{f}$-state distribution is well separated.
Figure~\ref{fig:3}(c) shows the assignment matrix.
We define the false-flag rate $\varepsilon_{\mathrm{FF}}$ as the probability that a prepared computational state is assigned to $\{\ket{f}\}$, averaged the initial states~$\ket{g}$ and $\ket{e}$.
We also define the undetected-leakage rate $\varepsilon_{\mathrm{UL}}$ as the probability that the prepared $\ket{f}$ state is assigned to $\gesubspace$.
The assignment matrix gives $\varepsilon_{\mathrm{FF}} = 2.3(3)\%$ and $\varepsilon_{\mathrm{UL}} = 3.5(2)\%$.
We define the leakage-detection fidelity as 
\begin{equation}
    \mathcal{F}_{\mathrm{d}} = 1 - \frac{\varepsilon_{\mathrm{FF}} + \varepsilon_{\mathrm{UL}}}{2}
\end{equation}
and obtain $\mathcal{F}_{\mathrm{d}} = 97.1(3)\%$.

To quantify these errors, we adapt the error-decomposition framework of Ref.~\citenum{sunada_fast_2022} to our binary classification~(see Sec.~III of the Supplemental Material for details).
The framework decomposes the total detection error into separation errors $\varepsilon_{\mathrm{sep}}$ arising from finite signal-to-noise ratio and flip errors $\varepsilon_{\mathrm{flip}}$ arising from state transitions during the detection.
We obtain $\varepsilon_{\mathrm{sep}} \approx 2\%$ in both directions.
The flip contributions to the error rates are $0.2\%$ for $\varepsilon_{\mathrm{FF}}$ and $1.6\%$ for $\varepsilon_{\mathrm{UL}}$.
The larger flip contribution to $\varepsilon_{\mathrm{UL}}$ reflects the decay of $\ket{f}$ during the detection.

\begin{figure}
    \centering
    \includegraphics{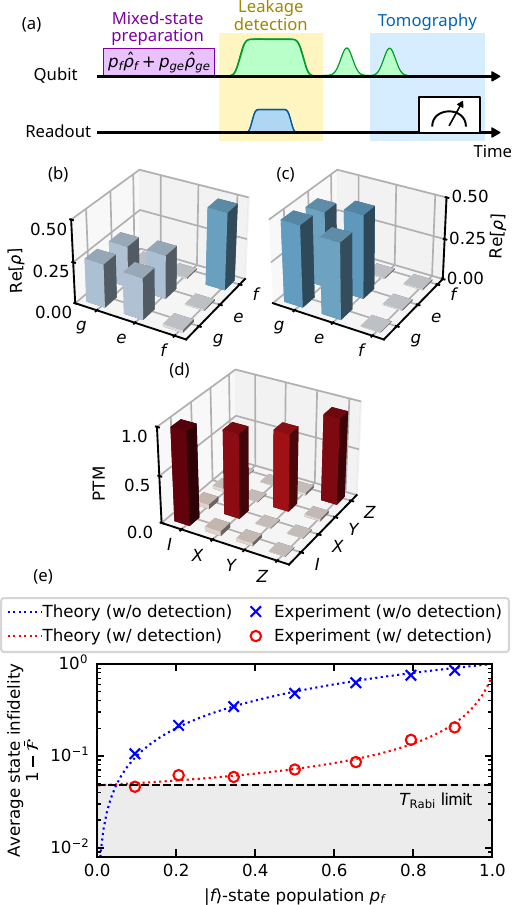}
    \caption{(a)~Pulse sequence for the leakage detection and post-selection experiment. (b)~Real part of the density matrix for the prepared state $(\ketbra{f}{f} + \ketbra{+}{+})/2$. (c)~Real part of the density matrix after leakage detection and conditioning on the $\gesubspace$ outcome, starting from the same initial state shown in~(b). (d)~Pauli transfer matrix~$R$ of the leakage detection for the $\ket{g}$--$\ket{e}$ subspace, measured at $p_f=0.5$. (e)~Average state infidelity~$1-\bar{\mathcal{F}}$ as a function of $\ket{f}$-state population~$p_f$. Blue dotted curve and crosses are theoretical prediction and the experimental results without leakage detection. Red dotted curve and circles are with leakage detection and conditioning on the no-leakage outcome. The black dashed line indicates the $T_\mathrm{Rabi}$ limit.}
    \label{fig:4}
\end{figure}
Finally, we demonstrate the leakage detection on a state containing both computational and leakage populations.
The projective character of the detection on the qutrit subspace is examined in Sec.~VII of the Supplemental Material.
We prepare a mixed state $p_f \op{\rho}_f+p_{ge}\op{\rho}_{ge}$, where $p_f\,(p_{ge})$ is the population of $\ket{f}$ state ($\ket{g}$--$\ket{e}$ subspace), $\op{\rho}_f=\ketbra{f}{f}$, and $\op{\rho}_{ge}$ is the density operator within the $\ket{g}$--$\ket{e}$ subspace.
The incoherent mixture is realized using a resonator-assisted Raman transition~\cite{zeytinoglu_microwave-induced_2015, pechal_microwave-controlled_2014} (see Sec.~IV of the Supplemental Material).
As $\op{\rho}_{ge}$, we prepare the six cardinal states of $\ketbra{g}{g}$, $\ketbra{e}{e}$, $\ketbra{\pm}{\pm}$, and $\ketbra{\!\pm i}{\pm i}$ where $\ket{\pm}=(\ket{g}\pm\ket{e})/\sqrt{2}$ and $\ket{\pm i}=(\ket{g}\pm i\ket{e})/\sqrt{2}$. 
Figure~\ref{fig:4}(b) shows the density matrix reconstructed via quantum state tomography for a target of $p_f=0.5$ and $\op{\rho}_{ge}=\ketbra{+}{+}$.
The measured $\ket{f}$-state population is $p_f=0.471(6)$, and the state fidelity of the $\ket{g}$--$\ket{e}$ subspace is 97.6(5)\%, confirming the successful preparation of the target mixed state.

The Rabi drive rotates the $\ket{g}$--$\ket{e}$ subspace during the detection, and this rotation must be undone to recover the input state.
It cannot be made trivial by truncating the drive at a multiple of $2\pi$ because the ac Stark shift from the readout photons tilts the rotating axis toward the $z$ axis.
The strong drive of $\OR/2\pi=186$~MHz also limits the timing resolution.
We therefore allow an arbitrary drive duration and undo the residual rotation with a recovery gate $\op{R}_z(\phi)\op{R}_x(\theta)\op{R}_z(\lambda)$ applied immediately after the drive, where $\op{R}_{z(x)}(\varphi)$ denotes a rotation around the $z\,(x)$-axis by an angle $\varphi$.
The three parameters $(\phi,\,\theta,\,\lambda)$ are calibrated to maximize the state fidelity~(see Sec.~V of the Supplemental Material for the details).

After the mixed-state preparation, we apply the leakage detection protocol with the recovery gate under the same detection conditions as in Fig.~\ref{fig:3}, and record the classification outcome for each shot~[Fig.~\ref{fig:4}(a)].
Shots classified as $\{\ket{f}\}$ in the leakage detection are discarded, and quantum state tomography is performed on the remaining shots.
Figure~\ref{fig:4}(c) shows the reconstructed density matrix after the detection for $p_f=0.5$ and $\op{\rho}_{ge}=\ketbra{+}{+}$.
The $\ket{f}$-state population is strongly suppressed and the $\ket{g}$--$\ket{e}$ subspace retains the $\ketbra{+}{+}$ structure.
The average state fidelity~$\bar{\mathcal{F}}$ over the six states after the detection reaches 92.9(5)\% for $p_f=0.5$.
Figure~\ref{fig:4}(d) shows the Pauli transfer matrix~$R$ for the qubit subspace, yielding an average fidelity of the conditional-detection process $\mathcal{F}_{\mathrm{proc}}=(\mathrm{Tr}[R]+2)/6=93.5(5)\%$~\cite{chow_universal_2012}.

Figure~\ref{fig:4}(e) plots the average state infidelity~$1-\bar{\mathcal{F}}$ as a function of the mixing weight~$p_f$.
Without leakage detection, the infidelity grows linearly with $p_f$, as the $\ket{f}$ population directly degrades the fidelity.
With heralded leakage detection, the infidelity is substantially reduced for the range of $p_f$ measured in this experiment~($0.10\le p_f\le 0.90$).
For $p_f\to 0$, the infidelity approaches the limit of approximately 4.7\%. This value is determined by the residual dephasing under the Rabi drive, which is consistent with the $T_{\mathrm{Rabi}}$ measured in Fig.~\ref{fig:2}.
On the other hand, as $p_f$ increases, the infidelity rises due to the detection errors.
We derive an analytical expression for $1-\bar{\mathcal{F}}$ by modeling the post-selected state including these error contributions (see Sec.~VI of the Supplemental Material).
The resulting prediction agrees well with the experimental data, validating the detection protocol and the error-budget analysis.

In summary, we have demonstrated leakage detection of a fixed-frequency transmon while suppressing measurement-induced back-action on the computational subspace.
A detuned Rabi drive on the $\ket{g}$--$\ket{e}$ transition during dispersive readout makes the computational states indistinguishable and compensates for the residual dephasing that arises from the finite anharmonicity.
Single-shot binary classification of $\gesubspace$ and $\{\ket{f}\}$ achieves a leakage-detection fidelity of 97.1(3)\% within an 80\nobreakdash-ns window, with false-flag and undetected-leakage rates of 2.3(3)\% and 3.5(2)\%, respectively.
Quantum state tomography on a mixture of computational and leakage populations confirms that the post-detection state retains an average fidelity of $92.9(5)\%$ for $p_f=0.5$.

As indicated by its agreement with the $T_{\mathrm{Rabi}}$ limit in Fig.~\ref{fig:4}(e), the conditional infidelity is limited by measurement-induced dephasing during the detection.
Since the photon-number noise scales as $S_{nn}(\OR)\propto\nbar\kappa$, reducing the effective resonator linewidth $\kappa_{\mathrm{eff}}/2\pi\approx53$~MHz suppresses the back-action at the cost of a slower ring-up~\cite{sete_quantum_2015}.
An error model calibrated to the measured error budget gives an optimal linewidth $\kappa_{\mathrm{eff}}\approx|\chi|$, improving the state infidelity approximately threefold and reducing the false-flag rate below 1\%~(see Sec.~VIII of the Supplemental Material).
Narrowing the Purcell filter bandwidth suppresses the noise beyond single-pole scaling, giving a back-action of approximately $10^{-3}$ at a separation error of $10^{-3}$ with present coherence times, or $2\times10^{-4}$ with an order-of-magnitude improvement in $T_1$~(see Sec.~IX of the Supplemental Material).
At this level of performance, the present leakage-detection protocol could be useful in quantum error correction, where the leakage population per cycle is expected to be small.
It could also be exploited in erasure-qubit architectures~\cite{kubica_erasure_2023,levine_demonstrating_2024,chou_demonstrating_2023}.

\begin{acknowledgments}
We thank S.~Shirai and A.~Hesse for fruitful discussions and A.~Fedorov for kindly answering our questions on quantum rifling.
This work was supported in part by the University of Tokyo Forefront Physics and Mathematics Program to Drive Transformation (FoPM), a World-leading Innovative Graduate Study (WINGS) Program, the Ministry of Education, Culture, Sports, Science and Technology (MEXT) Quantum Leap Flagship Program (Q-LEAP) (Grant No.\ JPMXS0118068682), the JSPS Grant-in-Aid for Scientific Research (KAKENHI) (Grant No.\ JP22H04937), the JST CREST (Grant No.\ JPMJCR23I4), and the Japan Science and Technology Agency (JST) Adopting Sustainable Partnerships for Innovative Research Ecosystem (ASPIRE)~(Grant Number JPMJAP2513).
\end{acknowledgments}

\clearpage
\FloatBarrier

\clearpage
\onecolumngrid
\begin{center}
\textbf{\large Supplemental Material for ``Heralded Leakage Detection with Preserved Computational-State Coherence in a Fixed-Frequency Transmon''}
\end{center}
\vspace{1em}
\twocolumngrid
\setcounter{section}{0}
\setcounter{equation}{0}
\setcounter{figure}{0}
\setcounter{table}{0}

\setcounter{secnumdepth}{3}
\renewcommand{\thefigure}{S\arabic{figure}}
\renewcommand{\theequation}{S\arabic{equation}}

\section{Experimental setup}\label{app:setups}
\begin{figure}[t]
    \centering
    \includegraphics{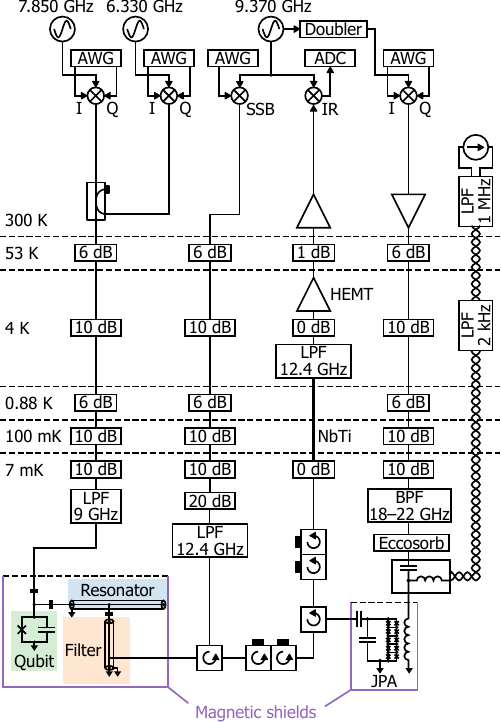}
    \caption{Measurement setup used in the experiment. AWG, arbitrary waveform generator; ADC, analog-to-digital converter; SSB, single sideband mixer; IR, image reject mixer; LPF, low-pass filter; HEMT, high-electron-mobility transistor; BPF, band-pass filter; JPA, Josephson parametric amplifier.}
    \label{fig:setup}
\end{figure}
\begin{figure}
    \centering
    \includegraphics{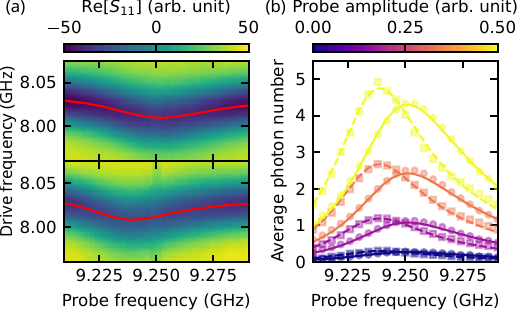}
    \caption{Characterization of the readout resonator via the ac-Stark-shift method~\cite{sank_system_2025-SM}. (a)~Real part of the reflection coefficient $\mathrm{Re}[S_{11}]$ as a function of the probe frequency and the $\pi_{ge}$-pulse frequency. The upper (lower) panel corresponds to the qubit initialized in $\ket{g}\,(\ket{e})$. The red curves show the ac-Stark-shifted $\ket{g}$--$\ket{e}$ transition frequency at each probe frequency. (b)~Mean intra-cavity photon number as a function of the probe frequency for different probe amplitudes. Circles~(squares) show the photon numbers $\nbar_g\,(\nbar_e)$ extracted from the ac-Stark-shift measurement with the qubit in $\ket{g}\,(\ket{e})$. Solid~(dashed) curves are the fits to Eq.~\eqref{eq:photon_num}.}
    \label{fig:6}
\end{figure}
Figure~\ref{fig:setup} shows the measurement setup used in this work.
Three microwave drive lines are connected to the device. They drive the qubit $\ket{g}$--$\ket{e}$ and $\ket{e}$--$\ket{f}$ transitions, the readout resonator, and the resonator-assisted Raman transition for active reset~\cite{magnard_fast_2018-SM} and mixed-state preparation (Sec.~\ref{app:mixed_state_prep}). 
The device, which was also used in Refs.~\citenum{miyamura_generation_2025-SM}~and~\citenum{wang_generation_2026-SM}, consists of a fixed-frequency transmon coupled to two independent resonator--filter systems.
In this work, we use the resonator--filter system that was not employed in those references.
The qubit coherence times are listed in Table~\ref{tab:coherence_times}.
The device parameters are summarized in Table~\ref{tab:parameters} and are extracted from ac-Stark-shift measurements~\cite{sank_system_2025-SM} as follows.
A probe tone is applied to the resonator, and after 500~ns a $\pi_{ge}$ pulse is applied while the probe remains on.
A subsequent dispersive readout detects whether the qubit state has flipped or not.
From the ac-Stark-shifted transition frequency as a function of the probe frequency, we extract $\omega_{\mathrm{r}}$, $\omega_{\mathrm{f}}$, $J$, $\kappa_{\mathrm{f}}$, and $\chi$~[Fig.~\ref{fig:6}(a)].

The mean photon number $\nbar=(\nbar_g+\nbar_e)/2$ is used throughout this work.
Here $\nbar_{g(e)}$ is the steady-state intra-cavity photon number with the qubit in $\ket{g}\,(\ket{e})$, given by
\begin{equation}\label{eq:photon_num}
    \nbar_{g(e)}(\omega_{\mathrm{p}}) = \left|\frac{\xi J}{(\omega_{\mathrm{r},g(e)}-\omega_{\mathrm{p}})(\omega_{\mathrm{f}}-\omega_{\mathrm{p}} - i\kappa_{\mathrm{f}}/2)-J^2}\right|^2,
\end{equation}
where $\omega_{\mathrm{p}}$ is the probe frequency, $\xi$ is the probe amplitude, $\omega_{\mathrm{r},g}=\omega_{\mathrm{r}}$, and $\omega_{\mathrm{r},e}=\omega_{\mathrm{r}}+\chi$.
Figure~\ref{fig:6}(b) shows $\nbar_{g}$ and $\nbar_{e}$ as a function of the probe frequency for different probe amplitudes.

\begin{table}[b]
    \caption{Coherence times of the qubit.}
    \centering
    \begin{ruledtabular}
    \begin{tabular}{lrr}
    $\ket{g}$--$\ket{e}$ energy-relaxation time & $T_{1, ge}$ & $26.7\pm7.1\mathrm{\,\mu s}$\\
    $\ket{g}$--$\ket{e}$ Ramsey dephasing time & $T_{2, ge}^*$ & $8.0\pm1.2\mathrm{\,\mu s}$\\
    $\ket{g}$--$\ket{e}$ echo dephasing time & $T_{2, ge}^{\mathrm{e}}$ & $18.1\pm3.7\mathrm{\,\mu s}$\\
    $\ket{e}$--$\ket{f}$ energy-relaxation time & $T_{1,ef}$ & $16.3\pm3.0\mathrm{\,\mu s}$\\
    \end{tabular}
    \end{ruledtabular}
    \label{tab:coherence_times}
\end{table}
\begin{table}[tbh]
    \caption{Parameters of the system.}
    \centering
    \begin{ruledtabular}
    \begin{tabular}{lrr}
    Qubit frequency & $\omega_{eg}/2\pi$ & $8.0355\mathrm{\,GHz}$\\
    Qubit anharmonicity & $\alpha/2\pi$ & $-318.0\mathrm{\,MHz}$\\
    Resonator frequency & $\omega_{\mathrm{r}}/2\pi$ & $9.2631\mathrm{\,GHz}$\\
    Filter frequency & $\omega_{\mathrm{f}}/2\pi$ & $9.286\mathrm{\,GHz}$\\
    Resonator--filter coupling strength & $J/2\pi$ & $50.8\mathrm{\,MHz}$\\
    External coupling of the filter & $\kappa_{\mathrm{f}}/2\pi$ & $184\mathrm{\,MHz}$\\
    Qubit--resonator full dispersive shift & $\chi/2\pi$ & $-10.3\mathrm{\,MHz}$\\
    \end{tabular}        
    \end{ruledtabular}
    \label{tab:parameters}
\end{table}

\section{Theory}\label{sec:theory}
In this section, we provide the theory of the leakage-detection back-action suppression. 
Here we consider a single resonator dispersively coupled to a transmon qubit for simplicity, but the theory is applicable to the case where a Purcell filter is present between the resonator and the feed line~(see Sec.~\ref{app:twopole}).
\subsection{System Hamiltonian}\label{sec:hamiltonian}
We model the transmon as a weakly anharmonic oscillator.
In the rotating-wave approximation, the system Hamiltonian $\op{\mathcal{H}}$ is
\begin{equation}\label{eq:H_system}
  \op{\mathcal{H}}/\hbar = \omega_{\mathrm{r}}\op{a}^\dag\op{a} + \omega_{eg}\op{b}^\dag\op{b}
  + \frac{\alpha}{2}\op{b}^\dag\op{b}(\op{b}^\dag\op{b}-1)
  + \chi\op{a}^\dag\op{a}\op{b}^\dag\op{b},
\end{equation}
where $\op{a}$~($\op{b}$) is the resonator (transmon) annihilation operator, $\omega_{\mathrm{r}}$~($\omega_{eg}$) the resonator (qubit) frequency, $\alpha < 0$ the transmon anharmonicity, and $\chi$ the dispersive shift per photon, defined such that the resonator frequency shifts by $\chi$ when the transmon is excited from $\ket{g}$ to $\ket{e}$.

\subsection{Photon-number noise spectrum}\label{sec:psd}
When the resonator is driven by a coherent probe with mean intra-cavity photon number $\nbar$, the photon-number fluctuations $\dn(t) = \op{a}^\dag\op{a} - \nbar$ exhibit the autocorrelation~\cite{gambetta_qubit-photon_2006-SM,clerk_introduction_2010-SM}
\begin{equation}\label{eq:dn_corr}
  \avg{\dn(\tau)\dn(0)} = \nbar e^{-\kappa|\tau|/2},
\end{equation}
reflecting shot noise of a coherent state filtered by a resonator of linewidth $\kappa$.
The corresponding noise spectral density is
\begin{equation}\label{eq:Snn}
  S_{nn}(\omega)
  = \int_{-\infty}^{\infty}\!d\tau
    \avg{\dn(\tau)\dn(0)}\,e^{-i\omega\tau}
  = \frac{\nbar\kappa}{\omega^2 + (\kappa/2)^2},
\end{equation}
$S_{nn}$ behaves as a Lorentzian with half-width at half-maximum $\kappa/2$.

In the absence of any qubit drive, each intra-cavity photon shifts the qubit frequency by $\chi$, so that photon-number fluctuations produce a stochastic frequency shift $\delta\omega_{eg}(t) = \chi\dn(t)$.
In the Markov limit ($t \gg 2/\kappa$), the resulting measurement-induced dephasing rate is determined by the zero-frequency component of the qubit-frequency noise spectrum~\cite{devoret_circuit_2014-SM}
\begin{equation}\label{eq:Gamma_m}
  \Gm = \frac{1}{2}\,\chi^2\,S_{nn}(0) = \frac{2\chi^2\nbar}{\kappa},
\end{equation}
which is the standard expression of measurement-induced dephasing~\cite{gambetta_qubit-photon_2006-SM}.

% ======================================================================
\subsection{Introducing a Rabi drive with the two-level picture}\label{sec:rifling_2level}
% ======================================================================
We introduce a Rabi drive of strength~$\OR$ and frequency~$\omega_{\mathrm{d}}$ during readout.
Restricting the Hamiltonian to the computational subspace~$\{\ket{g},\ket{e}\}$ and working in the frame rotating at $\omega_{\mathrm{d}}$, the Hamiltonian reads
\begin{equation}\label{eq:H_rot_2level}
  \hat{H}_{\mathrm{rot}} =  \frac{\OR}{2}\left(\ketbra{g}{e}+\ketbra{e}{g}\right) + \chi\dn(t)\ketbra{e}{e},
\end{equation}
where we have dropped the ac-Stark-shift term for simplicity.
The eigenstates of the drive Hamiltonian are $\ket{\pm}=(\ket{g}\pm\ket{e})/\sqrt{2}$, separated by $\OR$.
Expressing the noise operator in the dressed basis,
\begin{equation}\label{eq:ee_dressed}
 \chi\dn(t) \ketbra{e}{e}
  = \frac{\chi\dn(t)}{2}\left(I_{ge} - \ketbra{+}{-} - \ketbra{-}{+}\right),
\end{equation}
where $I_{ge}=\ketbra{g}{g}+\ketbra{e}{e}$.
The noise acts as a purely transverse perturbation in the dressed basis, driving transitions between $\ket{+}$ and $\ket{-}$ separated by frequency $\OR$. 
It is therefore sampled at $\omega = \OR$ rather than at $\omega = 0$~\cite{devoret_circuit_2014-SM}.

The matrix element of the noise coupling between the two 
dressed states is $\chi/2$ [Eq.~\eqref{eq:ee_dressed}].
By Fermi's golden rule, the transition rate from 
$\ket{+}$ to $\ket{-}$ is 
$(\chi/2)^2\,S_{nn}(\OR)$, where the noise is sampled 
at the dressed-state splitting $\OR$.
Since $S_{nn}(\OR)=S_{nn}(-\OR)$, the reverse transition 
proceeds at the same rate, and the longitudinal 
relaxation rate is the sum of both,
\begin{equation}
  \Gonerho^{\mathrm{meas}}
  = 2\left(\frac{\chi}{2}\right)^{\!2} S_{nn}(\OR)
  = \frac{\chi^2\,\nbar\kappa}
    {2\bigl[\OR^2 + (\kappa/2)^2\bigr]}.
  \label{eq:G1rho_meas}
\end{equation}
The transverse relaxation rate follows from Bloch--Redfield 
theory~\cite{yan_rotating-frame_2013-SM,
smirnov_decoherence_2003-SM} as
\begin{equation}
  \Gtworho^{\mathrm{meas}}
  = \frac{1}{2}\,\Gonerho^{\mathrm{meas}}.
  \label{eq:GRabi_meas}
\end{equation}
Including intrinsic energy relaxation ($\Gamma_1 = 1/T_1$), the longitudinal and transverse relaxation rates $\Gonerho,\,\Gtworho=\Grabi$ are~\cite{yan_rotating-frame_2013-SM,smirnov_decoherence_2003-SM}
\begin{align}
  \Gonerho &= \frac{1}{2}\,\Gamma_1 + \Gonerho^{\mathrm{meas}}, \label{eq:Gamma1rho}
  \\
  \Gtworho&= \frac{3}{4}\,\Gamma_1 + \Gtworho^{\mathrm{meas}}. \label{eq:GRabi}
\end{align}
The suppression ratio for measurement-induced noise relative to the undriven case is
\begin{equation}\label{eq:suppression}
  \frac{S_{nn}(\OR)}{S_{nn}(0)}
  = \frac{(\kappa/2)^2}{\OR^2 + (\kappa/2)^2},
\end{equation}
which becomes effective when $\OR \gg \kappa$~\cite{szombati_quantum_2020-SM}.

% ======================================================================
\subsection{Three-level effects and the clock condition}
\label{sec:clock}
% ======================================================================

The analysis above treated the transmon as a simple two-level system.
In practice, the second excited state $\ket{f}$ modifies the noise structure of the dressed qubit.

In the frame rotating at the drive frequency $\omega_\mathrm{d}$, the transmon Hamiltonian truncated to $\{\ket{g},\ket{e},\ket{f}\}$ under a Rabi drive reads
\begin{equation}\label{eq:H_3level}
  \op{\mathcal{H}}_{\mathrm{rot}} =
  \begin{pmatrix}
    0 & \OR/2 & 0 \\
    \OR/2 & -\delta & \sqrt{2}\,\OR/2 \\
    0 & \sqrt{2}\,\OR/2 & -2\delta+\alpha
  \end{pmatrix},
\end{equation}
where $\delta \equiv \omega_\mathrm{d} - \omega_{eg}$ is the drive detuning from the $\ket{g}$--$\ket{e}$ transition frequency. The noise Hamiltonian in this truncated space is
\begin{equation}
    \op{V}(t) = \chi\dn
  \begin{pmatrix}
    0 & 0 & 0 \\
    0 & 1 & 0 \\
    0 & 0 & 2
  \end{pmatrix}.
\end{equation}

\begin{figure}
    \centering
    \includegraphics{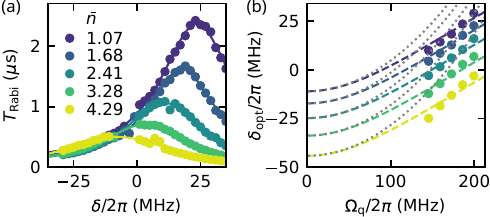}
    \caption{(a)~$T_{\mathrm{Rabi}}$ as a function of drive detuning $\delta$ for different average photon numbers $\nbar$ at $\OR/2\pi=186\mathrm{\,MHz}$. Circles are experimental data, and solid curves are the Lorentzian fits. (b)~Optimal detuning $\delta_\mathrm{opt}$ as a function of drive strength $\OR$ at different $\nbar$. Circles are experimental values extracted from the peak positions of the Lorentzian fits. Dashed curves are the theoretical prediction obtained by numerical diagonalization of a six-level transmon Hamiltonian with the ac Stark shift $\chi\nbar$ included. Gray dotted curves are the leading-order approximation $\delta=-\OR^2/2\alpha + \chi\nbar$.}
    \label{fig:clock}
\end{figure}

When $\ket{f}$ is included, the dressed computational states $\ket{\tilde{0}}$~and $\ket{\tilde{1}}$ acquire $\ket{f}$ admixtures through the $\ket{e}$--$\ket{f}$ coupling.
Projecting the transmon excitation-number operator onto the dressed computational subspace yields
\begin{equation}\label{eq:noise_decomp}
    \op{P}_{\mathrm{D}}\op{b}^\dag\op{b}\op{P}_{\mathrm{D}}
  = \bar{N}\op{P}_{\mathrm{D}} + \frac{\Delta N}{2}\hat{\sigma}^\mathrm{D}_z
  + V_\perp\hat{\sigma}^\mathrm{D}_x,
\end{equation}
where $\op{P}_{\mathrm{D}} = \ketbra{\tilde 0}{\tilde 0}+\ketbra{\tilde 1}{\tilde 1}$ is the projector onto the dressed computational subspace, $\hat{\sigma}^\mathrm{D}_z=\ketbra{\tilde 1}{\tilde 1}-\ketbra{\tilde 0}{\tilde 0}$, and $\hat{\sigma}^\mathrm{D}_x=\ketbra{\tilde 0}{\tilde 1} + \ketbra{\tilde 1}{\tilde 0}$.
The coefficients are $\bar{N}=(N_{\tilde 1} + N_{\tilde 0})/2$, $\Delta N = N_{\tilde 1} - N_{\tilde 0}$, and $V_\perp = \bra{\tilde{0}}\op{b}^\dag\op{b}\ket{\tilde{1}}$, with $N_{\tilde j}=\bra{\tilde{j}}\op{b}^\dag\op{b}\ket{\tilde{j}}$.
The phases of $\ket{\tilde 0}$ and $\ket{\tilde 1}$ are chosen such that $V_\perp$ is real.
The full noise Hamiltonian in this subspace is $\chi\dn(t)\,\op{P}_{\mathrm{D}}\op{b}^\dag\op{b}\op{P}_{\mathrm{D}}$.
In the two-level limit~($\alpha\to-\infty$), one has $\Delta N = 0$ and $V_\perp = 1/2$, recovering the purely transverse noise of Sec.~\ref{sec:rifling_2level}.
For finite anharmonicity, the virtual $\ket{e}$--$\ket{f}$ coupling affects $\ket{e}$ while $\ket{g}$ remains unaffected, which produces $\Delta N\neq 0$.
A nonzero $\Delta N$ generates a longitudinal noise in the dressed basis, which samples $S_{nn}(0)$ and therefore evades the drive-induced back-action suppression.
The resulting dressed-basis pure dephasing rate~$\Gphirho^{\mathrm{meas}}$ is
\begin{equation}\label{eq:Gphirho_meas}
  \Gphirho^{\mathrm{meas}}
  = \frac{1}{2}\,(\chi\Delta N)^2\,S_{nn}(0)
  = \frac{2\chi^2(\Delta N)^2\nbar}{\kappa}.
\end{equation}
To leading order in $|\OR/\alpha|$, a perturbative treatment in the $\ket{e}$--$\ket{f}$ coupling at $\delta=0$ yields
\begin{equation}
    \Delta N\Bigl|_{\delta=0} = -\frac{\OR}{2\alpha} + O\left((\OR/\alpha)^2\right).
\end{equation}
This residual dephasing can be compensated by introducing a small drive detuning $\delta$.
The detuning tilts the quantization axis in the $\ket{g}$--$\ket{e}$ subspace and results in 
\begin{equation}
    \Delta N \approx -\frac{\delta}{\sqrt{\OR^2 + \delta^2}} + \Delta N\Bigl|_{\delta=0} .
\end{equation}
Setting $\Delta N=0$ and using $|\OR|\gg|\delta|$ yields the optimal detuning, presented in Eq.~\eqref{eq:delta_clock_main}.
Since $\alpha<0$ for transmons, the compensation requires a positive detuning, consistent with the experimental observation in the main text.
To leading order in $\OR/\alpha$, this condition agrees with the spin-locking theory for transmons~\cite{zuk_robust_2024-SM}.
The ac Stark shift from the readout photons displaces the effective qubit frequency by $\chi\nbar$.
Since $\chi<0$, the optimal drive detuning from the bare $\ket{g}$--$\ket{e}$ frequency $\omega_{eg}$ decreases with increasing probe power.
Figure~\ref{fig:clock}(a) shows $T_{\mathrm{Rabi}}$ for different resonator photon numbers $\nbar$ at $\OR/2\pi=186\mathrm{\,MHz}$.
As $\nbar$ increases, the peak position shifts to lower detunings, reflecting the ac Stark shift.

At the experimental operating point, higher-order corrections to Eq.~\eqref{eq:delta_clock_main} become significant.
Figure~\ref{fig:clock}(b) shows the optimal drive detuning as a function of $\OR$ for different $\nbar$.
The experimental values agree with the theoretical predictions obtained by numerical diagonalization of a six-level transmon Hamiltonian, whereas the leading-order expression in Eq.~\eqref{eq:delta_clock_main} overestimates the optimal detuning at large $\OR$.

% ======================================================================
\subsection{Connection to the leakage-detection protocol}
\label{sec:fstate_theory}
% ======================================================================
We now connect the dressed-basis relaxation rates to the back-action on the bare-basis density matrix during the leakage detection.
Consider an input state $\ket{\psi} = \alpha\ket{g} + \beta\ket{e}$ in the computational subspace.
In the dressed basis, this state reads
\begin{equation}
    \ket{\psi} = \frac{(\alpha+\beta)\ket{+} + (\alpha-\beta)\ket{-}}{\sqrt{2}}.
\end{equation}
The off-diagonal element of $\op{\rho}_\psi = \ketbra{\psi}{\psi}$ in the bare basis, $\rho_{ge}=\bra{g}\op{\rho}_\psi\ket{e}$, can be decomposed using the dressed-basis matrix elements $\rho_{\mu\nu}=\bra{\mu}\rho_{\psi}\ket{\nu}\;(\mu,\,\nu \in \{+,\,-\})$ as 
\begin{equation}\label{eq:rho_ge_decomp}
  \rho_{ge} = \frac{1}{2}(\rho_{++}-\rho_{--}) - \frac{1}{2}(\rho_{+-}-\rho_{-+}).
\end{equation}
The first term is the population difference in the dressed basis and decays at the rate $\Gonerho$.
Since $\rho_{-+}=\rho_{+-}^*$, the second term equals $-i\,\mathrm{Im}[\rho_{+-}]$, which oscillates at $\OR$ and decays at the rate $\Gtworho$.
Similarly, the diagonal elements are
\begin{align}
    \rho_{gg} &= \frac{1}{2} + \mathrm{Re}[\rho_{+-}], \\
    \rho_{ee} &= \frac{1}{2} - \mathrm{Re}[\rho_{+-}],
\end{align}
which are governed solely by $\Gtworho$.

\subsection{Rabi-oscillation fitting}\label{app:fitting}
The Rabi-oscillation traces such as the one shown in Fig.~\ref{fig:2}(b) are fitted to the function~\cite{kosugi_theory_2005-SM}
\begin{equation}\label{eq:rabi_fit}
    f(t)=A\,e^{-t/T_\mathrm{Rabi}}\cos(\Omega_\mathrm{R} t + \phi) + B\,e^{-t/T_{1\rho}} + C,
\end{equation}
where $\Omega_\mathrm{R}=\sqrt{\OR^2+\delta^2}$ is the generalized Rabi frequency.
The decay times $T_{\mathrm{Rabi}}=1/\Gtworho$ and $T_{1\rho}=1/\Gonerho$ are the transverse and longitudinal relaxation times of the dressed states in Eqs.~\eqref{eq:Gamma1rho} and \eqref{eq:GRabi}.
The free parameters are the amplitudes~$A$, $B$, and~$C$, the phase~$\phi$, the frequency~$\Omega_{\mathrm{R}}$ and the two decay times.

The second term describes the non-oscillating component of the signal.
For a detuned drive, the rotation axis tilts out of the equatorial plane of the Bloch sphere.
The Bloch vector then acquires a component along the tilted axis.
This component does not oscillate and decays at the longitudinal rate $\Gonerho$.
To leading order, its amplitude $B$ is proportional to the detuning $\delta$~\cite{kosugi_theory_2005-SM}.
The finite $B$ makes the oscillations asymmetric about the offset $C$, as observed in Fig.~\ref{fig:2}(b).
We extract $T_{\mathrm{Rabi}}$ from the fit and plot it in Fig.~\ref{fig:2}(c).

% \subsection{Spurious mode in the $T_{\mathrm{Rabi}}$ measurement}
\subsection{Anomalous Rabi oscillations}\label{app:beating}
\begin{figure}
    \centering
    \includegraphics{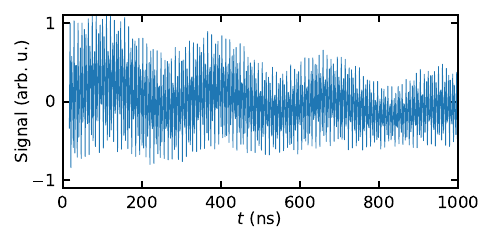}
    \caption{Anomalous Rabi oscillation at $\nbar=2.41$, $\OR/2\pi=199$~MHz, and $\delta/2\pi=8$~MHz.}
    \label{fig:beating}
\end{figure}

In measuring $T_{\mathrm{Rabi}}$, we observed a beating modulation of the Rabi oscillation at certain drive conditions, which prevented a reliable extraction of $T_{\mathrm{Rabi}}$ at those points~[Fig.~\ref{fig:beating}].
The affected drive detuning shifted toward positive values as the drive strength $\OR$ was increased, while it depended little on the probe power.
The weak probe-power dependence indicates that the anomaly is not caused by the readout photons, and we attribute it to a spurious mode near the qubit frequency.
For the strongest drive of $\OR/2\pi=199$~MHz, the affected detuning fell close to the value that maximizes $T_{\mathrm{Rabi}}$, which is the reason why we operate at $\OR/2\pi=186$~MHz instead. The affected data points are omitted from Fig.~\ref{fig:2}(c).

\begin{figure}
    \centering
    \includegraphics{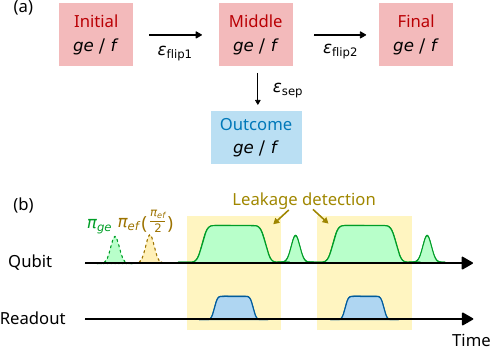}
    \caption{(a)~Schematic of a single leakage-detection process.
    (b)~Pulse sequence of the two consecutive detections used for the error analysis.}
    \label{fig:readout}
\end{figure}

\begin{table}[b]
    \caption{Error budget of the leakage detection.}
    \centering
    \begin{ruledtabular}
    \begin{tabular}{lcc}
    Separation error& $\varepsilon_{\mathrm{sep}}^{ge\to f}$ & 2.1(2)\%\\
                    & $\varepsilon_{\mathrm{sep}}^{f\to ge}$ & 1.9(1)\%\\
    Early flip error& $\varepsilon_{\mathrm{flip1}}^{ge\to f}$ & 0.2(2)\%\\
                    & $\varepsilon_{\mathrm{flip1}}^{f\to ge}$ & 1.6(1)\%\\
    Late flip error & $\varepsilon_{\mathrm{flip2}}^{ge\to f}$ & 0.1(1)\%\\
                    & $\varepsilon_{\mathrm{flip2}}^{f\to ge}$ & 1.5(1)\%\\
                    
    \end{tabular}
    \end{ruledtabular}
    \label{tab:error_budget}
\end{table}
\section{Detection-error analysis}\label{app:error_analysis}
We analyze the readout errors using the framework of Ref.~\citenum{sunada_fast_2022-SM}, which decomposes a single readout into separation errors $\varepsilon_\mathrm{sep}$ due to finite signal-to-noise ratio and flip errors $\varepsilon_{\mathrm{flip}}$ from state transitions during the detection.
The flip errors are further classified into early flip errors $\varepsilon_\mathrm{flip1}$ occurring before the classification decision, and late flip errors $\varepsilon_\mathrm{flip2}$ occurring after it~[Fig.~\ref{fig:readout}(a)].
Since the Rabi drive merges $\ket{g}$ and $\ket{e}$ into a single resonator response, our readout yields a binary outcome ``$\{\ket{g},\,\ket{e}\}$'' or ``$\{\ket{f}\}$'' and the two-level error model in Ref.~\citenum{sunada_fast_2022-SM} applies directly with the identification $\ket{g} \to \{\ket{g},\,\ket{e}\}$ and $\ket{e} \to \{\ket{f}\}$.

We extract $\varepsilon_\mathrm{sep}$ and $\varepsilon_\mathrm{flip}$ from two consecutive leakage detections following Ref.~\citenum{sunada_fast_2022-SM}~[Fig.~\ref{fig:readout}(b)].
Three initial states with distinct ratios of $\{\ket{f}\}$ to $\gesubspace$ outcomes in the first detection are prepared. 
We use $\ket{g}$ prepared by an active reset, $\ket{f}$ via successive $\pi_{ge}$ and $\pi_{ef}$ pulses, and $(\ket{e}+\ket{f})/\sqrt{2}$ via a $\pi_{ge}$ pulse followed by a $\pi_{ef}/2$ pulse.
To separate $\varepsilon_\mathrm{flip1}$ and $\varepsilon_\mathrm{flip2}$, we use the single-shot readout results in Fig.~\ref{fig:3}.
In a single readout, the late flip error $\varepsilon_\mathrm{flip2}$ does not affect the measurement outcome.
Assuming negligible state-preparation error, the misassignment probability in a single detection equals $\varepsilon_\mathrm{sep}+\varepsilon_\mathrm{flip1}$, because a late flip does not affect the outcome.
Since $\varepsilon_\mathrm{sep}$ is determined independently from the two consecutive detections, we obtain the early flip error as the difference between the misassignment probability and the separation error.
The late flip error then follows from $\varepsilon_\mathrm{flip2}=\varepsilon_{\mathrm{flip}}-\varepsilon_{\mathrm{flip1}}$.
The error rates in the main text therefore decompose as 
\begin{align}
    \varepsilon_{\mathrm{FF}} = \varepsilon_{\mathrm{sep}}^{ge\to f} + \varepsilon_{\mathrm{flip1}}^{ge\to f}, \label{eq:H_FF}\\
    \varepsilon_{\mathrm{UL}} = \varepsilon_{\mathrm{sep}}^{f\to ge} + \varepsilon_{\mathrm{flip1}}^{f\to ge},\label{eq:H_UL}
\end{align}
and the flip contributions quoted in the main text correspond to $\varepsilon_{\mathrm{flip1}}$.
The extracted error probabilities are summarized in Table~\ref{tab:error_budget}.

\begin{table*}[t]
\caption{All sixteen pathways through the detection.}
\centering
\begin{ruledtabular}
\begin{tabular}{ccccccccc}
\# & Initial & flip1 & Middle & classification & flip2 & Final & Declared (Outcome) & $\mathcal{F}$ \\
\hline
1  & $ge$ & $1\!-\!\varepsilon_{\mathrm{flip1}}^{ge\to f}$ & $ge$ & $1\!-\!\varepsilon_{\mathrm{sep}}^{ge\to f}$ & $1\!-\!\varepsilon_{\mathrm{flip2}}^{ge\to f}$ & $ge$ & $ge$ & $\mathcal{F}_{\mathrm{Rabi}}$ \\
2  & $ge$ & $1\!-\!\varepsilon_{\mathrm{flip1}}^{ge\to f}$ & $ge$ & $1\!-\!\varepsilon_{\mathrm{sep}}^{ge\to f}$ & $\varepsilon_{\mathrm{flip2}}^{ge\to f}$       & $f$  & $ge$ & $0$ \\
3  & $ge$ & $1\!-\!\varepsilon_{\mathrm{flip1}}^{ge\to f}$ & $ge$ & $\varepsilon_{\mathrm{sep}}^{ge\to f}$       & $1\!-\!\varepsilon_{\mathrm{flip2}}^{ge\to f}$ & $ge$ & $f$  & --- \\
4  & $ge$ & $1\!-\!\varepsilon_{\mathrm{flip1}}^{ge\to f}$ & $ge$ & $\varepsilon_{\mathrm{sep}}^{ge\to f}$       & $\varepsilon_{\mathrm{flip2}}^{ge\to f}$       & $f$  & $f$  & --- \\
5  & $ge$ & $\varepsilon_{\mathrm{flip1}}^{ge\to f}$       & $f$  & $1\!-\!\varepsilon_{\mathrm{sep}}^{f\to ge}$ & $1\!-\!\varepsilon_{\mathrm{flip2}}^{f\to ge}$ & $f$  & $f$  & --- \\
6  & $ge$ & $\varepsilon_{\mathrm{flip1}}^{ge\to f}$       & $f$  & $1\!-\!\varepsilon_{\mathrm{sep}}^{f\to ge}$ & $\varepsilon_{\mathrm{flip2}}^{f\to ge}$       & $ge$ & $f$  & --- \\
7  & $ge$ & $\varepsilon_{\mathrm{flip1}}^{ge\to f}$       & $f$  & $\varepsilon_{\mathrm{sep}}^{f\to ge}$       & $1\!-\!\varepsilon_{\mathrm{flip2}}^{f\to ge}$ & $f$  & $ge$ & $0$ \\
8  & $ge$ & $\varepsilon_{\mathrm{flip1}}^{ge\to f}$       & $f$  & $\varepsilon_{\mathrm{sep}}^{f\to ge}$       & $\varepsilon_{\mathrm{flip2}}^{f\to ge}$       & $ge$ & $ge$ & $1/2$ \\
\hline
9  & $f$   & $1\!-\!\varepsilon_{\mathrm{flip1}}^{f\to ge}$ & $f$  & $1\!-\!\varepsilon_{\mathrm{sep}}^{f\to ge}$ & $1\!-\!\varepsilon_{\mathrm{flip2}}^{f\to ge}$ & $f$  & $f$  & --- \\
10 & $f$   & $1\!-\!\varepsilon_{\mathrm{flip1}}^{f\to ge}$ & $f$  & $1\!-\!\varepsilon_{\mathrm{sep}}^{f\to ge}$ & $\varepsilon_{\mathrm{flip2}}^{f\to ge}$       & $ge$ & $f$  & --- \\
11 & $f$   & $1\!-\!\varepsilon_{\mathrm{flip1}}^{f\to ge}$ & $f$  & $\varepsilon_{\mathrm{sep}}^{f\to ge}$       & $1\!-\!\varepsilon_{\mathrm{flip2}}^{f\to ge}$ & $f$  & $ge$ & $0$ \\
12 & $f$   & $1\!-\!\varepsilon_{\mathrm{flip1}}^{f\to ge}$ & $f$  & $\varepsilon_{\mathrm{sep}}^{f\to ge}$       & $\varepsilon_{\mathrm{flip2}}^{f\to ge}$       & $ge$ & $ge$ & $1/2$ \\
13 & $f$   & $\varepsilon_{\mathrm{flip1}}^{f\to ge}$       & $ge$ & $1\!-\!\varepsilon_{\mathrm{sep}}^{ge\to f}$ & $1\!-\!\varepsilon_{\mathrm{flip2}}^{ge\to f}$ & $ge$ & $ge$ & $1/2$ \\
14 & $f$   & $\varepsilon_{\mathrm{flip1}}^{f\to ge}$       & $ge$ & $1\!-\!\varepsilon_{\mathrm{sep}}^{ge\to f}$ & $\varepsilon_{\mathrm{flip2}}^{ge\to f}$       & $f$  & $ge$ & $0$ \\
15 & $f$   & $\varepsilon_{\mathrm{flip1}}^{f\to ge}$       & $ge$ & $\varepsilon_{\mathrm{sep}}^{ge\to f}$       & $1\!-\!\varepsilon_{\mathrm{flip2}}^{ge\to f}$ & $ge$ & $f$  & --- \\
16 & $f$   & $\varepsilon_{\mathrm{flip1}}^{f\to ge}$       & $ge$ & $\varepsilon_{\mathrm{sep}}^{ge\to f}$       & $\varepsilon_{\mathrm{flip2}}^{ge\to f}$       & $f$  & $f$  & --- \\
\end{tabular}
\end{ruledtabular}
\label{tab:pathways}
\end{table*}

\section{Mixed-state preparation}\label{app:mixed_state_prep}
\begin{figure}[t]
    \centering
    \includegraphics{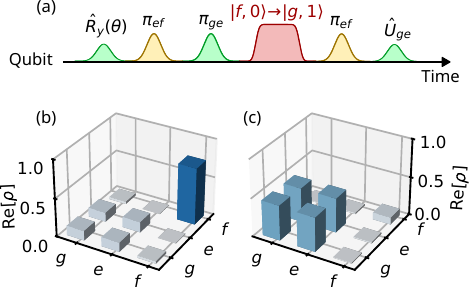}
    \caption{(a)~Pulse sequence to prepare the mixed state $p_f\op{\rho}_f + p_{ge}\op{\rho}_{ge}$. The initial rotation angle $\theta$ sets the weight $p_f=\cos^2(\theta/2)$, and the final gate $\op{U}_{ge}$ sets the computational-subspace component. (b)(c)~Real parts of the reconstructed density matrices for (b)~$\theta=0.3\pi\,(p_f\approx 0.79)$ and (c)~$\theta=0.8\pi\,(p_f\approx 0.10)$, with $\op{U}_{ge}=\op{R}_y(\pi/2)$.}
    \label{fig:mixed_state}
\end{figure}
In this section we explain how we prepare mixed states of the form $p_f\ketbra{f}{f} + p_{ge}\op{\rho}_{ge}$.
The pulse sequence for the mixed-state preparation is shown in Fig.~\ref{fig:mixed_state}(a).
Starting from $\ket{g}$, we apply a rotation $\op{R}_{y}(\theta)$ of angle $\theta$ in the $\ket{g}$--$\ket{e}$ subspace, followed by a $\pi_{ef}$ pulse and a $\pi_{ge}$ pulse, which together generate the pure superposition $\cos(\theta/2)\ket{e}+\sin(\theta/2)\ket{f}$.
A resonator-assisted Raman transition on the $\ket{f,0}$--$\ket{g,1}$ transition~\cite{zeytinoglu_microwave-induced_2015-SM, pechal_microwave-controlled_2014-SM}, where the numbers in the ket denote the Fock state of the resonator, then maps $\ket{f,0}$ to $\ket{g,1}$ while leaving $\ket{e,0}$ unchanged, producing the qubit--resonator entangled state $\cos(\theta/2)\ket{e,0}+\sin(\theta/2)\ket{g,1}$.
After the resonator photon escapes through the readout port, tracing out the photonic degree of freedom reduces the qubit state to the diagonal ensemble $\cos^2(\theta/2)\ketbra{e}{e}+\sin^2(\theta/2)\ketbra{g}{g}$. 
A subsequent $\pi_{ef}$ pulse excites the $\ket{e}$ population to $\ket{f}$, and a final gate $\op{U}_{ge}$ rotates the remaining $\ket{g}$ population to the target state within the $\ket{g}$--$\ket{e}$ subspace without affecting the $\ket{f}$ component. 
The resulting density operator takes the form $p_f\ketbra{f}{f} + p_{ge}\op{\rho}_{ge}$ with $p_f=\cos^2(\theta/2)$ so that the angle~$\theta$ continuously tunes the weight of the $\ket{f}$ component. 
Examples of the prepared density matrices are shown in Figs.~\ref{fig:mixed_state}(b) and (c).

\section{Calibration of recovery gate}\label{app:recovery_gate}
The three parameters~($\phi,\,\theta,\,\lambda$) of the recovery gate $\op{R}_z(\phi)\op{R}_x(\theta)\op{R}_z(\lambda)$ are calibrated so that the combined action of the detection sequence and the recovery gate approximates an identity operation on the computational subspace.
To quantify the calibration, we prepare each of the six cardinal states~$\{\ket{g},\,\ket{e},\,\ket{\pm},\,\ket{\pm i}\}$ by applying a unitary gate $\op{U}_k$ to $\ket{g}$.
We then apply the detection and recovery sequence, followed by the inverse gate~$\op{U}_k^\dag$.
If the detection and recovery together act as the identity, the qubit returns to $\ket{g}$.
The figure of merit is the ground-state population~$P_g^{(k)}$ averaged over the six states, $\bar{P}_g = \sum_{k=1}^6 P_g^{(k)}/6$.
The ${R}_z$ rotations are realized as virtual~$Z$ gates, which update the reference frame of subsequent pulses without applying any physical microwave pulse~\cite{mckay_efficient_2017-SM}.
We optimize $\bar{P}_g$ using Bayesian optimization with a Gaussian-process surrogate model.
Starting from 10 uniformly sampled initial points, the optimization converges within approximately 150 evaluations to $\bar{P}_g\approx 0.93$.

\section{State fidelity after leakage detection}\label{app:state_fidelity}
We evaluate the average state fidelity of the $\ket{g}$--$\ket{e}$ subspace conditioned on the detection outcome declaring ``no leakage''.
We consider an input state of the form $\rho_{\mathrm{in}} = p_{ge}\ketbra{\psi}{\psi} + p_f\ketbra{f}{f}$, where $\ket{\psi}$ is an arbitrary state in the $\ket{g}$--$\ket{e}$ subspace and $p_f = 1 - p_{ge}$.

Table~\ref{tab:pathways} enumerates all sixteen pathways through the detection. Each pathway's weight is the product of the initial population and the three probability factors in the table.
Of these, eight pathways (rows 1, 2, 7, 8, 11, 12, 13, 14) yield a ``$ge$'' declaration.
We assign a fidelity to each declared-$ge$ pathway as follows.
Row~1, where no flip or misclassification occurs, retains fidelity
\begin{equation}
\mathcal{F}_{\mathrm{Rabi}} 
= \frac{1 + e^{-\bar{\Gamma}\,t_{\mathrm{R}}}}{2}
\label{eq:F_Rabi}
\end{equation}
to the input, where $\bar{\Gamma}=(2\Gtworho+\Gonerho)/3$ is the six-state-averaged relaxation rate and $t_{\mathrm{R}}$ is the detection window.
Rows~2, 7, 11, and 14, whose final states are $\ket{f}$, contribute zero fidelity.
Rows~8, 12, and 13, which pass through $\ket{f}$ at any stage and end in $\gesubspace$, contribute fidelity $1/2$, as the passage through $\ket{f}$ destroys the correlation with the input.

Summing over all pathways, the average state fidelity conditioned on a ``$ge$'' declaration is
\begin{equation}
\bar{\mathcal{F}} = \frac{\mathcal{N}}{\mathcal{D}},
\label{eq:F_state_exact}
\end{equation}
where
\begin{align*}
\mathcal{D} &= p_{ge}\!\left[(1-a_1)(1-b_1) + a_1 b_2\right] \\
  & \qquad \qquad \qquad+ p_f\!\left[(1-a_2) b_2 + a_2(1-b_1)\right], 
\end{align*}
\begin{align*}
\mathcal{N} &= p_{ge}(1-a_1)(1-b_1)(1-c_1)\,\mathcal{F}_{\mathrm{Rabi}} \\
  &\qquad\qquad + \frac{1}{2}\!\left[p_{ge} a_1 b_2 c_2 + p_f(1-a_2) b_2 c_2 
  \right. \\
  & \qquad\qquad\qquad\qquad\left.+ p_f a_2(1-b_1)(1-c_1)\right], 
\end{align*}
with abbreviations $a_i = \varepsilon_\mathrm{flip1}$,
$b_i = \varepsilon_\mathrm{sep}$, $c_i = \varepsilon_\mathrm{flip2}$ for
$i=1\;(ge\to f)$ and $i=2\;(f\to ge)$.
The denominator $\mathcal{D}$ sums all paths that yield a ``$ge$'' outcome, while the numerator $\mathcal{N}$ weights each path by the fidelity of its post-measurement state to the input.
Expanding to leading order in the error probabilities,
\begin{equation}
1 - \bar{{\mathcal{F}}} \approx
  \frac{\bar{\Gamma}\, t_{\mathrm{R}}}{2}
  + \varepsilon_\mathrm{flip2}^{ge \to f}
  + \frac{p_f}{p_{ge}}\!\left(\varepsilon_\mathrm{sep}^{f \to ge} + \frac{1}{2}\varepsilon_\mathrm{flip1}^{f \to ge}\right),
\label{eq:infidelity}
\end{equation}
where $\varepsilon_\mathrm{sep}^{ge \to f}$ and $\varepsilon_{\mathrm{flip1}}^{ge \to f}$ do not appear at the first order because they reduce both $\mathcal{N}$ and $\mathcal{D}$ by the same fraction.
The first term represents coherence loss during the detection.
The second term accounts for post-detection leakage from the $\ket{g}$--$\ket{e}$ subspace to $\ket{f}$.
The third term captures contamination from $\ket{f}$-state events that enter the $\gesubspace$ ensemble, weighted by the relative population $p_f/p_{ge}$.

\section{Leakage detection as a projection measurement}\label{app:projection}
\begin{figure}[t]
    \centering
    \includegraphics{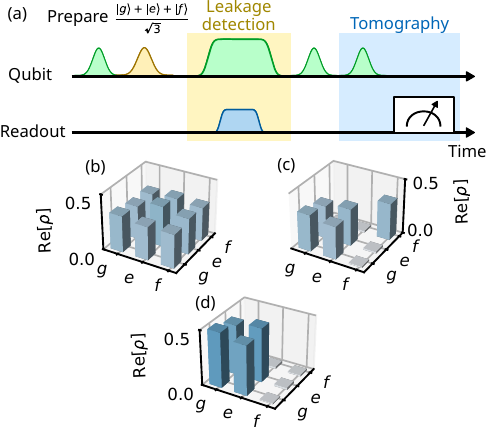}
    \caption{(a)~Pulse sequence for the leakage-detection experiment on a coherent qutrit superposition. (b)~Real part of the reconstructed density matrix of the prepared input state $(\ket{g}+\ket{e}+\ket{f})/\sqrt{3}$. (c)~Real part of the reconstructed density matrix after the leakage detection and recovery gate, without conditioning on the detection outcome. (d)~Same as (c) but post-selected on the $\gesubspace$ outcome.}
    \label{fig:gef}
\end{figure}
The leakage-detection protocol is expected to act as a projective measurement with POVM elements $\{\ketbra{f}{f},\,I-\ketbra{f}{f}\}$.
To verify this, we prepare the coherent superposition $(\ket{g}+\ket{e}+\ket{f})/\sqrt{3}$~[Fig.~\ref{fig:gef}(b)], apply the detection with the recovery gate, and perform quantum state tomography~[Fig.~\ref{fig:gef}(a)].
Figure~\ref{fig:gef}(c) shows the reconstructed density matrix without conditioning on the detection outcome.
The diagonal populations remain at $1/3$, and the $\ket{g}$--$\ket{e}$ coherence is largely retained, while the $\ket{g}$--$\ket{f}$ and $\ket{e}$--$\ket{f}$ coherences are strongly suppressed.
Figure~\ref{fig:gef}(d) shows the density matrix post-selected on the $\gesubspace$ outcome.
The $\ket{f}$ population is removed and the computational subspace retains the $\ketbra{+}{+}$ structure, with a state fidelity of 94.7(8)\% to $\ketbra{+}{+}$.

\section{Projected performance versus resonator linewidth}\label{app:kappa}

In the present device, a band-pass Purcell filter sets the effective resonator linewidth to $\kappa_{\mathrm{eff}}/2\pi\approx 53$~MHz.
Here we estimate how the detection performance scales with $\kappa$ by building a model of the four detection errors, and then varying $\kappa$ while optimizing the probe parameters at each value.

We use $\ket{f}$ as the leakage state and treat the detection as a binary discrimination between two resonator responses.
Under a sufficiently strong Rabi drive, the computational subspace pulls the resonator by the time-averaged shift $(0+\chi)/2=\chi/2$.
The $\ket{f}$-state pull is approximately $2\chi$.
For a step-function probe pulse of amplitude $\xi$ at detuning $\Delta_{\mathrm{p}}=\omega_{\mathrm{p}}-\omega_{\mathrm{r}}$ from the bare resonator frequency, the steady-state intra-cavity amplitude is $\alpha_w = \xi/[(\Delta_{\mathrm{p}} - w) - i\kappa/2]$, where $w=\chi/2$ for the computational subspace and $w=2\chi$ for $\ket{f}$.
We denote the two amplitudes by $\alpha_{\mathrm{c}}$ and $\alpha_{\mathrm{f}}$, and the corresponding photon numbers by $\nbar_{\mathrm{c}}=|\alpha_{\mathrm{c}}|^2$ and $\nbar_{\mathrm{f}}=|\alpha_{\mathrm{f}}|^2$.
The quantity $\nbar_{\mathrm{c}}$ corresponds to the mean photon number $\nbar$ used in the main text.
The intra-cavity field builds up as $\alpha_w(1 - e^{-\kappa t/2})$ after the probe is switched on~\cite{gambetta_protocols_2007-SM}.
The signal for discrimination accumulates over the effective integration time
\begin{equation}
\tau_{\mathrm{eff}}
 = \int_0^{t_{\mathrm{R}}}\!\left(1 - e^{-\kappa t/2}\right)^2 dt,
\label{eq:taueff}
\end{equation}
where $t_{\mathrm{R}}$ is the detection window.
Photon-induced errors continue to accumulate during the ring-down after the window closes. 
We write $\tau_{\mathrm{tot}}$ for the total photon exposure 
including this ring-down contribution.
\begin{figure}[t]
    \centering
    \includegraphics{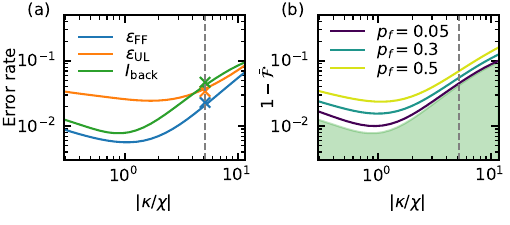}
    \caption{(a)~Optimized error rates as a function of $|\kappa/\chi|$. Blue, the false-flag rate~[Eq.~\eqref{eq:H_FF}]. Orange, the undetected-leakage rate~[Eq.~\eqref{eq:H_UL}]. Green, back-action contribution~[Eq.~\eqref{eq:iback}]. The dashed line marks the present device~($\kappa/2\pi\approx53$~MHz) and the crosses show the corresponding measured values.
    (b)~Conditional state infidelity~$1-\bar{\mathcal{F}}$ for three values of $p_f$. Green shaded area indicates the back-action limit, which corresponds to the green curve in (a).}
    \label{fig:kappa}
\end{figure}

The separation error from finite signal-to-noise ratio is~\cite{gambetta_protocols_2007-SM}
\begin{equation}
    \varepsilon_{\mathrm{sep}} = \frac{1}{2}\,
\mathrm{erfc}\!\left(\frac{\sqrt{S}}{2\sqrt{2}}\right),
\end{equation}
with
\begin{equation}
    S = 4\eta\kappa
|\alpha_{\mathrm{c}} - \alpha_{\mathrm{f}}|^2\tau_{\mathrm{eff}},
\end{equation}
where $\eta$ is the measurement efficiency~\cite{gambetta_protocols_2007-SM, walter_rapid_2017-SM}.
We model the separation error to be symmetric, i.e., $\varepsilon_{\mathrm{sep}}^{ge\to f}
=\varepsilon_{\mathrm{sep}}^{f\to ge}
\equiv\varepsilon_{\mathrm{sep}}$.

The back-action on the computational subspace contributes to the conditional infidelity as
\begin{equation}
I_{\mathrm{back}}
 = \frac{1}{2}\!\left(\frac{2}{3}\Gamma_1 t_{\mathrm{R}}
 + \frac{4}{3}C_{\mathrm{back}}\,\Gtworho^{\mathrm{meas}}
   \,\tau_{\mathrm{tot}}\right),
\label{eq:iback}
\end{equation}
corresponding to the first term of Eq.~\eqref{eq:infidelity}.
The coefficients $2/3$ and $4/3$ arise from the six-state average over the dressed-basis decay rates $\Gonerho$ and $\Gtworho$ as described in Sec.~\ref{app:state_fidelity}.
The intrinsic relaxation is weighted by $t_{\mathrm{R}}$, whereas the measurement-induced dephasing follows the photon exposure $\tau_{\mathrm{tot}}$.
Here $\Gtworho^{\mathrm{meas}}$($=\Gonerho^{\mathrm{meas}}/2$) is evaluated from Eqs.~\eqref{eq:G_meas_main} and \eqref{eq:Snn_lorentz}, and $C_{\mathrm{back}}$ absorbs the deviation of the actual noise spectrum from the single-Lorentzian form
and the excess dephasing of unidentified origin at the present operating point~(see Sec.~\ref{app:twopole}).

The flip errors combine intrinsic relaxation with measurement-induced transitions,
\begin{align}
p_{f\to ge} &= \Gamma_{1,ef}\,t_{\mathrm{R}}
 + \gamma_{\mathrm{rel}}\,\nbar_{\mathrm{f}}\,
   \tau_{\mathrm{tot}},
\label{eq:pfge}\\
p_{ge\to f} &= p_0
 + \gamma_{\mathrm{leak}}\,\nbar_{\mathrm{c}}\,
   \tau_{\mathrm{tot}},
\label{eq:pgef}
\end{align}
where $\gamma_{\mathrm{rel}}$ and $\gamma_{\mathrm{leak}}$ are the per-photon relaxation and leakage rates, respectively, and $p_0$ is the probe-independent contribution of the pulse ramps.
Under the strong Rabi drive, the dressed computational states acquire a finite $\ket{f}$ admixture~(see Sec.~\ref{sec:clock}). Thus, the photon-number noise $\chi\delta\op{n}\,\op{b}^\dag\op{b}$ develops off-diagonal matrix elements between the dressed computational subspace and $\ket{f}$, which drive $ge\leftrightarrow f$ transitions at a rate proportional to $S_{nn}\propto\nbar$~[Eq.~\eqref{eq:Snn}].
We adopt this linear scaling as a phenomenological model for the photon-dependent component of the flip rates.
The intrinsic rates $\Gamma_1$, $\Gamma_{1,ef}$, and $p_0$ are measured independently and carry no free coefficient.
Assuming that flips occur uniformly in time within the window, we divide them equally between early and late errors. Flips during the ring-down are counted as late errors.

The composite error rates follow from these four contributions.
The conditional infidelity is given by Eq.~\eqref{eq:infidelity},
\begin{equation}
1-\bar{\mathcal{F}} = I_{\mathrm{back}} 
 + \varepsilon_{\mathrm{flip2}}^{ge\to f}
 + \frac{p_f}{p_{ge}}\!\left(\varepsilon_{\mathrm{sep}}
 + \frac{1}{2}\varepsilon_{\mathrm{flip1}}^{f\to ge}
 \right).
\label{eq:H_infid}
\end{equation}
The false-flag and undetected-leakage rates follow the decomposition in Eqs.~\eqref{eq:H_FF} and~\eqref{eq:H_UL}.

The four coefficients ($\eta$, $C_{\mathrm{back}}$, $\gamma_{\mathrm{rel}}$, $\gamma_{\mathrm{leak}}$) are determined at the present operating condition with $\kappa/2\pi=53$~MHz and $t_{\mathrm{R}}=80$~ns.
We set the probe power such that $\nbar_{\mathrm{c}}$ matches the measured value of $2.28$ at the detection operation point~(Fig.~\ref{fig:3}), yielding $\nbar_{\mathrm{f}}=2.5$ from the response model~[Eq.~\eqref{eq:photon_num}].
$C_{\mathrm{back}}$ is set so that $I_{\mathrm{back}}$ reproduces the measured $T_{\mathrm{Rabi}}=1.07\mathrm{\,\mu s}$.
$\eta$ is determined so that $\varepsilon_{\mathrm{sep}}$ matches the average of the measured separation errors $(\varepsilon_{\mathrm{sep}}^{ge\to f}+\varepsilon_{\mathrm{sep}}^{f\to ge})/2\approx 2.0\%$, resulting in $\eta=0.25$.
$\gamma_{\mathrm{rel}}$ and $\gamma_{\mathrm{leak}}$ are set so that the flip probabilities reproduce Table~\ref{tab:error_budget}.
The probe-independent transition probability~$p_0=0.2\%$ is obtained from a master-equation simulation of transmon--resonator--filter system under the Rabi drive without the readout probe.

For each $\kappa$, we numerically optimize the probe detuning, the probe power, and the detection window to minimize the conditional infidelity in Eq.~\eqref{eq:H_infid} at $p_f=0.5$.
The photon numbers are capped at $\nbar_{\mathrm{f}}\le 12$ and $\nbar_{\mathrm{c}}\le 6$ to remain within the dispersive regime.
Figure~\ref{fig:kappa}(a) shows the three error rates at the optimized operating point for each $\kappa$.
All three rates reach a minimum near $\kappa\approx|\chi|$, close to the linewidth that maximizes the signal-to-noise ratio in standard dispersive readout.
At this optimum, the false-flag rate and back-action contribution both fall below 1\%, while the undetected-leakage rate remains at 2--3\%, dominated by~$\varepsilon_{\mathrm{flip1}}^{f\to ge}$.
Figure~\ref{fig:kappa}(b) shows the conditional infidelity at three values of $p_f$.
At $p_f=0.05$, the infidelity is reduced to approximately 1\%.
For smaller $p_f$, the conditional infidelity approaches the back-action contribution $I_{\mathrm{back}}$.

\section{Photon-number noise in the presence of the Purcell filter}
\label{app:twopole}

We extend the noise model of Sec.~\ref{sec:psd} to include the Purcell filter and estimate the back-action in the regime relevant to quantum error correction.
In that regime, we suppose that the leakage population satisfies $p_f \lesssim 10^{-2}$, the third term of Eq.~\eqref{eq:infidelity} is negligible, and the conditional infidelity is dominated by the back-action contribution $I_{\mathrm{back}}$.
In this appendix, we examine whether $I_{\mathrm{back}}$ can be reduced below $10^{-3}$ and which parameters are required.
As an important result, the noise spectral density falls off as $\OR^{-4}$ when the drive strength $\OR$ is much larger than the filter linewidth $\kappa_{\mathrm{f}}$.
The flip contribution $\varepsilon_{\mathrm{flip2}}^{ge\to f}$ transfers population to the leakage states, which are flagged in the subsequent detection cycle and are therefore not counted as unheralded errors here.

\subsection{Noise spectrum density of a resonator--filter system}
Here, we calculate the noise spectrum density~$S_{nn}(\omega)$ of the resonator coupled to a Purcell filter.
We consider the case where we apply a probe at frequency~$\omega_{\mathrm{p}}$ onto the filter.
In the frame rotating at $\omega_{\mathrm{p}}$, the linearized Langevin equations of the resonator--filter system read
\begin{align}
\delta\dot{\hat{a}} &= -\left(i\Delta_{\mathrm{r}}
   + \frac{\kappa_{\mathrm{int}}}{2}\right)\delta\hat{a}
   - iJ\,\delta\hat{c} + \sqrt{\kappa_{\mathrm{int}}}\,\hat{a}_{\mathrm{in}},
\label{eq:langevin-a}\\
\delta\dot{\hat{c}} &= -\left(i\Delta_{\mathrm{f}}
   + \frac{\kappa_{\mathrm{f}}}{2}\right)\delta\hat{c}
   - iJ\,\delta\hat{a} + \sqrt{\kappa_{\mathrm{f}}}\,\hat{c}_{\mathrm{in}},
\label{eq:langevin-c}
\end{align}
where $\delta\hat{c}$ is the filter fluctuation operator, $\Delta_{\mathrm{r(f)}} = \omega_{\mathrm{r(f)}} - \omega_{\mathrm{p}}$ is the detuning between the filter~(resonator) frequency and the probe frequency, $\kappa_{\mathrm{int}}$ is the internal loss rate of the resonator, and
$\hat{a}_{\mathrm{in}}$ and $\hat{c}_{\mathrm{in}}$ are independent vacuum inputs.
Solving in the Fourier domain yields $\delta\hat{a}[\omega] = T_{\mathrm{int}}(\omega)\hat{a}_{\mathrm{in}}[\omega] + T_{\mathrm{ext}}(\omega)\hat{c}_{\mathrm{in}}[\omega]$,
with
\begin{align}
T_{\mathrm{ext}}(\omega) &= \frac{-iJ\sqrt{\kappa_{\mathrm{f}}}}{D(\omega)},
\quad
T_{\mathrm{int}}(\omega) = \frac{\sqrt{\kappa_{\mathrm{int}}}
  \left[\dfrac{\kappa_{\mathrm{f}}}{2} + i(\Delta_{\mathrm{f}} - \omega)\right]}
  {D(\omega)},
\nonumber\\
D(\omega) &= \left[\frac{\kappa_{\mathrm{int}}}{2}
  + i(\Delta_{\mathrm{r}} - \omega)\right]
  \left[\frac{\kappa_{\mathrm{f}}}{2} + i(\Delta_{\mathrm{f}} - \omega)\right]
  + J^2.
\label{eq:transfer}
\end{align}
Using the linearization
$\delta\hat{n} \approx \alpha^*\delta\hat{a}
+ \alpha\,\delta\hat{a}^{\dagger}$ with
$\nbar = |\alpha|^2$, the noise spectrum~[Eq.~\eqref{eq:Snn}] generalizes to
\begin{equation}
S_{nn}(\omega) = \frac{\nbar}{2}\sum_{s=\pm}
\left[\,|T_{\mathrm{ext}}(s\omega)|^2 + |T_{\mathrm{int}}(s\omega)|^2\,\right],
\label{eq:Snn2pole}
\end{equation}
where the sum over sidebands corresponds to the symmetrized rate in Eq.~\eqref{eq:G1rho_meas}.

\subsection{Suppression of measurement-induced noise}
The two roots of $D(\omega)=0$ correspond to a resonator-like mode and a filter-like mode.
For~$J^2 \ll \Delta_{\mathrm{fr}}^2 + (\kappa_{\mathrm{f}}/2)^2$, the resonator-like pole has the effective linewidth
\begin{equation}
\kappa_{\mathrm{eff}} = \kappa_{\mathrm{int}}
+ \frac{\kappa_{\mathrm{f}}\, J^2}{\Delta_{\mathrm{fr}}^2
+ (\kappa_{\mathrm{f}}/2)^2},
\label{eq:keff}
\end{equation}
where $\Delta_{\mathrm{fr}} = \omega_{\mathrm{f}} - \omega_{\mathrm{r}}$.
The parameters in Table~\ref{tab:parameters} give $\kappa_{\mathrm{eff}}/2\pi = 53$~MHz.
For noise frequencies within the filter bandwidth~($|\omega - \Delta_{\mathrm{f}}| \lesssim \kappa_{\mathrm{f}}/2$), Eq.~\eqref{eq:Snn2pole} reduces to the single-pole form~[Eq.~\eqref{eq:Snn_lorentz}] with $\kappa = \kappa_{\mathrm{eff}}$.
Outside the bandwidth, the vacuum noise from the feed line is suppressed twice, once by the filter response and once by the resonator response.
For~$|\omega - \Delta_{\mathrm{r}}| \gg \kappa_{\mathrm{eff}}/2$ and $|\omega - \Delta_{\mathrm{f}}| \gg \kappa_{\mathrm{f}}/2$, the external transfer function falls as
\begin{equation}
|T_{\mathrm{ext}}(\omega)|^2 \approx
\frac{J^2 \kappa_{\mathrm{f}}}
{(\omega - \Delta_{\mathrm{r}})^2\,(\omega - \Delta_{\mathrm{f}})^2}
\propto \omega^{-4},
\label{eq:tail}
\end{equation}
compared with $\omega^{-2}$ for a single resonator.
Neglecting~$\kappa_{\mathrm{int}}$, the suppression relative to the single-pole~$S_{nn}^\mathrm{1pole}(\omega)$ with the same $\kappa_{\mathrm{eff}}$ is, for~$\OR \gg \kappa_{\mathrm{f}}/2$ and $\OR \gg |\Delta_{\mathrm{fr}}|$,
\begin{equation}
\frac{S_{nn}(\OR)}{S_{nn}^{\mathrm{1pole}}(\OR)} \approx \frac{\Delta_{\mathrm{fr}}^2 + (\kappa_{\mathrm{f}}/2)^2}{\OR^2}.
\label{eq:suppression_ratio}
\end{equation}
In the present device, this factor is $0.26$.
For a resonant filter with $\omega_{\mathrm{f}} = \omega_{\mathrm{r}}$, Eq.~\eqref{eq:suppression_ratio} reduces to $(\kappa_{\mathrm{f}}/2\OR)^2$.
Narrowing the filter bandwidth~$\kappa_{\mathrm{f}}$ moves the noise sidebands deeper into the stop-band and strengthens the suppression~(Fig.~\ref{fig:appendixI}(a)).

Vacuum noise also enters through the internal loss of the resonator with rate~$\kappa_{\mathrm{int}}$.
This noise couples to the resonator directly and is suppressed only once.
The corresponding transfer function retains a single-pole tail, $|T_{\mathrm{int}}(\omega)|^2 \approx \kappa_{\mathrm{int}}/\omega^2$, setting the noise floor $S_{nn}(\OR) \gtrsim \nbar\kappa_{\mathrm{int}}/\OR^2$~(Fig.~\ref{fig:appendixI}(a)).
The filter suppression is effective when $\kappa_{\mathrm{int}} \ll \kappa_{\mathrm{eff}}\,[\Delta_{\mathrm{fr}}^2 + (\kappa_{\mathrm{f}}/2)^2]/\OR^2$.

\subsection{Back-action suppression in error-correction regime}
We now identify the conditions required to reduce $I_{\mathrm{back}}$ below $10^{-3}$.
The Rabi drive strength is fixed at the present value~$\OR/2\pi=186$~MHz, and the dispersive shift is swept.
Along with the sweeping, we set the parameters of a filter such that $\kappa_\mathrm{eff}=|\chi|$, the optimum found in Sec.~\ref{app:kappa}, and $\kappa_{\mathrm{f}} = 6\kappa_{\mathrm{eff}}$, which ensures the overdamped condition~$\kappa_{\mathrm{f}} > 4J$.
Also, we assume that the filter is on-resonant with the resonator~($\omega_\mathrm{f}=\omega_\mathrm{r}$).

We fix the separation error at $\varepsilon_{\mathrm{sep}} = 10^{-3}$ and
solve for the probe power that satisfies this target.
Here we assume a measurement efficiency $\eta = 0.25$ obtained in Sec.~\ref{app:kappa}.
The measurement-induced dephasing is evaluated from
Eqs.~\eqref{eq:G_meas_main} and \eqref{eq:Snn2pole} without the correction
factor $C_{\mathrm{back}}$, assuming that the excess noise discussed below
is removed.
The probe detuning and the detection window are optimized at each parameter set as in Sec.~\ref{app:kappa}.
Since the measurement-induced contribution is nearly independent of the
window at fixed $\varepsilon_{\mathrm{sep}}$, the optimal window is the
shortest one compatible with the photon-number caps.

\begin{figure}
\includegraphics{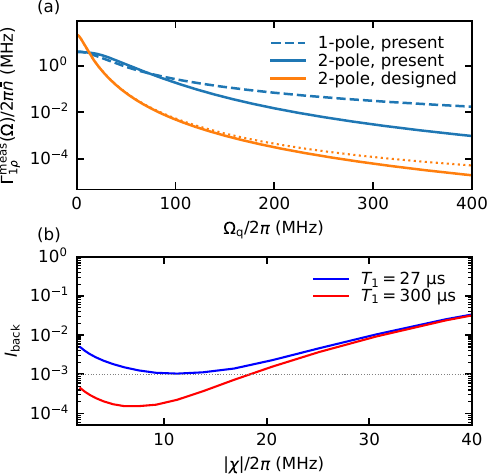}
\caption{(a)~Measurement-induced relaxation rate under a Rabi drive per intra-cavity photon~$\Gonerho^{\mathrm{meas}}/\nbar$ as a function of the Rabi-drive strength $\OR$. Blue dashed and solid curves show the single-pole approximation and the two-pole spectrum for the present filter with $\kappa_{\mathrm{eff}}/2\pi=53$~MHz and $\kappa_{\mathrm{f}}/2\pi=184$~MHz. The orange solid curve shows the modified design with $\kappa_{\mathrm{eff}}=|\chi|=10.3$~MHz and $\kappa_{\mathrm{f}}=6\kappa_{\mathrm{eff}}$. The orange dotted curve additionally includes an internal loss of the resonator of $\kappa_{\mathrm{int}}/2\pi=0.1$~MHz. 
(b)~Optimized back-action contribution $I_{\mathrm{back}}$ as a function of $|\chi|$, varied through $\chi$ at the fixed drive strength $\OR/2\pi=186$~MHz, for $T_1=27~\mathrm{\mu s}$~(blue) and $T_1=300~\mathrm{\mu s}$~(red).
}\label{fig:appendixI}
\end{figure}

Figure~\ref{fig:appendixI}(b) shows the optimized $I_{\mathrm{back}}$~[Eq.~\eqref{eq:iback}] as a function of $|\chi|$ for two values of $T_1$.
The two contributions of Eq.~\eqref{eq:iback} compete as $|\chi|$ is reduced.
A smaller $|\chi|$ narrows the filter bandwidth through the conditions $\kappa_{\mathrm{eff}}=|\chi|$ and $\kappa_{\mathrm{f}}=6\kappa_{\mathrm{eff}}$, and the fixed Rabi frequency lies deeper in the stop band.
The dephasing rate then scales as $\Gtworho^{\mathrm{meas}} \propto \chi^2 S_{nn}(\OR) \propto \nbar\,|\chi|^{5}/\OR^{4}$~[Eqs.~\eqref{eq:G_meas_main} and \eqref{eq:suppression_ratio}], whereas the photon exposure required for $\varepsilon_{\mathrm{sep}}=10^{-3}$ grows only as $\nbar\,\tau_{\mathrm{tot}} \propto 1/\kappa_{\mathrm{eff}}$.
The measurement-induced contribution $\Gtworho^{\mathrm{meas}}\,\tau_{\mathrm{tot}}$ therefore falls as $(|\chi|/\OR)^{4}$.
In contrast, the detection window lengthens as $1/\kappa_{\mathrm{eff}}$, and the intrinsic contribution grows as $1/(|\chi|\,T_1)$.
Minimizing the sum of the two contributions gives the optimum $|\chi|_{\mathrm{opt}} \propto T_1^{-1/5}$, at which $I_{\mathrm{back}}^{\mathrm{opt}} \propto T_1^{-4/5}$.
At the present values $|\chi|/2\pi = 10.3~\mathrm{MHz}$ and $T_1 = 27~\mathrm{\mu s}$, the back-action reaches $I_{\mathrm{back}} \approx 9\times10^{-4}$ with a 54-ns detection window, dominated by intrinsic relaxation.
Improving $T_1$ to $300~\mathrm{\mu s}$ reduces the back-action to approximately $2\times10^{-4}$.

The estimates above assume that the measurement-induced dephasing follows Eqs.~\eqref{eq:G_meas_main} and \eqref{eq:Snn2pole}.
However, the measured $\Gtworho^{\mathrm{meas}}$ exceeds this prediction by a factor of approximately three at the operating condition.
A similar excess appears without the Rabi drive~($\OR=0$).
We found that in the present device and setup, the measurement-induced dephasing rate under a weak probe exceeds Eq.~\eqref{eq:Gamma_m} by a factor of about $1.6$.
The origin of the excess has not been identified.
A plausible candidate is excess noise accompanying the probe tone, which acts on the qubit in the same way as the photon shot noise.
Removing this excess is a prerequisite for the estimates above.

\begingroup
\def\label#1{}%
\endgroup

\end{document}